\pdfoutput=1
\documentclass{aastex61}
\usepackage[T1]{fontenc}
\usepackage{ae,aecompl}
\usepackage{graphicx}
\usepackage[caption=false]{subfig}
\usepackage{epsf}
\usepackage{rotating}
\usepackage{color}
\usepackage{amsmath}

\usepackage{booktabs}
\usepackage{braket}
\usepackage{multirow}
\usepackage{natbib}
\begin{document}
\title{Exploring the brighter fatter effect with the Hyper Suprime-Cam}

\correspondingauthor{William R. Coulton}
\email{wcoulton@princeton.edu}
\author{William R. Coulton}
\affiliation{Joseph Henry Laboratories, Princeton University, Princeton, NJ 08544, USA}
\author{Robert Armstrong}
\affiliation{Department of Astrophysical Sciences, Princeton University,  Peyton Hall, Princeton, NJ 08544, USA}
\author{Kendrick M. Smith}
\affiliation{Perimeter Institute for Theoretical Physics, Waterloo, ON N2L 2Y5, Canada}
\author{Robert H. Lupton}
\affiliation{Department of Astrophysical Sciences, Princeton University,  Peyton Hall, Princeton, NJ 08544, USA}
\author{David N. Spergel}
\affiliation{Department of Astrophysical Sciences, Princeton University,  Peyton Hall, Princeton, NJ 08544, USA}
\affiliation{Center for Computational Astrophysics, Flatiron Institute,162 5th Avenue, 10010, New York, NY, USA}

\begin{abstract}
The brighter fatter effect has been postulated to arise due to the build up of a transverse electric field, produced as photo-charges accumulate in the pixels' potential wells. We investigate the brighter fatter effect in Hyper Suprime-Cam by examining flat fields and moments of stars. We observe deviations from the expected linear relation in the photon transfer curve, luminosity dependent correlations between pixels in flat field images and a luminosity dependent point spread function (PSF) in stellar observations. Under the key assumptions of translation invariance and Maxwell's equations in the quasi-static limit, we give a first-principles proof that the effect can be parametrized by a translationally invariant scalar kernel. We describe how this kernel can be estimated from flat fields and discuss how this kernel has been used to remove the brighter fatter distortions in Hyper Suprime-Cam images. We find that our correction restores the expected linear relation in the photon transfer curves and significantly reduces, but does not completely remove, the luminosity dependence of the PSF over a wide range of magnitudes. \end{abstract}
\section{Introduction}
 
When an astronomical image is taken there are many effects that can introduce variations across the image. These vary from camera based effects,such as varying quantum efficiencies across the CCDs, to optical system effects, such as vignetting.  Currently a range of calibration methods, such as the use of flat fields (nominally uniformly illuminated images), are used to control some of these effects.

The commonly calibrated effects have all left unchallenged the assumption that the counts in each pixel are independent of their neighbors and can be described by Poisson statistics. This assumption is known to be incorrect as it has been seen that the variance in a flat field image does not increase linearly with the mean, as is predicted by Poisson statistics. Instead the variance flattens off and covariances with neighboring pixels arise, such that when these covariances are summed the correct variance is attained \citep{downing2006}. A recent effect that has been seen in new thick-depleted CCDs is that the point spread function (PSF) has a dependence upon the intensity of the source \citep{Astier2014,Holland2014,Astier2013}. The effect is such that the brighter the PSF the wider it is, implying that the charge is spread out among the neighboring pixels. This effect appears to also violate the assumption that pixels are independent and poses potential problems for surveys interested in varying brightness effects, such as supernova searches. This will also be a problem for surveys that require precise information about the shape of objects, such as weak lensing surveys, as the measured shape will not represent the shape on the focal plane. The non-linear luminosity dependence of these effects mean that they cannot be accounted for by existing calibration methods, such as flat fielding.  

A physical source of this effect has been suggested by \citet{Astier2014}; they attribute both of the observed effects to build up of charge 
altering the effective pixel area. They suggest that as charge builds up in the potential wells a lateral electric field is generated, a field that gets larger as more charge accumulates in the potential well. This lateral field repels charges as they drift through the CCD bulk, from the point of photon conversion to the collecting gates. This repulsion means that the charges are deflected away from pixels that have accumulated large amounts of charge. If the point of conversion is close to the pixel boundary then the deflection by the accumulated charges can push the charge into the adjacent pixel. This idea has been supported by simulations \citep{rasmussen2014,rasmussen2014b}, where these effects are seen in detailed simulations of the large synoptic sky survey (LSST) CCDs \citep{lage2017}. 

Working in the limit of small pixels we develop a new theoretical treatment of the effect. With the assumption that we can treat the problem in the quasistatic limit, we form a set of equations describing how charge accumulates in the detector and how pixel correlation arise from the brighter fatter effect. We show that the effect is characterized by a translationally-invariant scalar kernel, which can be estimated solely from flat field images, an improvement upon the previous approach that is under-constrained from flat fields. Once reconstructed the kernel can be used to correct for the brighter fatter effect and we show how this has been used to significantly reduce this effect in images from the Hyper-Suprime-Cam.

With the increased use of thick-depleted CCDs and the consequences of this effect for science in future sky surveys, it is important to understand this and develop a correction. In section \ref{ch:HSCInst} we briefly review the Hyper-Suprime-Cam (HSC) instrument which was used to investigate this effect. In section \ref{ch:HSCPTC} we explore this effect in flat field data before moving on to present our proposed correction in section \ref{ch:correction}. We then present the results of applying our correction to flat field data in section \ref{ch:flatfield}, and to stars in section \ref{ch:stars} and then in section \ref{ch:lensingReq} compare how our correction performs with respect to the levels required by current and future weak lensing surveys .

\section{Hyper-Suprime Cam}
\label{ch:HSCInst}
The Hyper Suprime Cam (HSC) is an instrument that has been installed on the 8.2 meter Subaru telescope in Hawaii and 
has recently published first year science data \citep{HSCfirstyear}.  The instrument's optical system comprises  two main components: the wide field corrector (WFC) and the camera contained within the prime focus unit.  The wide field corrector consists of seven lenses which allows for a field of view of 1.5 degrees in diameter with an instrument point spread function with 80\% of the light contained in a diameter of less than 0.2" . The camera consists of a filter with an filter exchange unit (FEU) and a vacuum dewar containing and cooling 116 CCDS and their readouts providing 870 million pixels. Of the 116 CCDs, four are used for automatic guiding and eight are used for automatic focusing leaving 104 science CCDs. The instrument has a set of five broad band filters (\textit{grizy}) and a number of narrow band filters which will be exchanged by the FEU, though only six filters can be on the telescope at one time \citep{HSC2012,HSCcamera}.

The CCDs are made of N-type silicon, are backside illuminated and each CCD has four read out amplifiers. The CCDs are deep depleted and have a bulk thickness of 200 $\mu$m; the large bulk thickness increases the quantum efficiency of the device, particularly at the far red end of the spectrum, but at the cost of increased lateral charge diffusion. This large bulk makes these CCDs particularly sensitive to the brighter fatter effect \citep{HSCReview,HSCcamera}.
\section{HSC Flat Fields}
\label{ch:HSCPTC}
\begin{figure}
	\centering
	\includegraphics[width=0.5\textwidth]{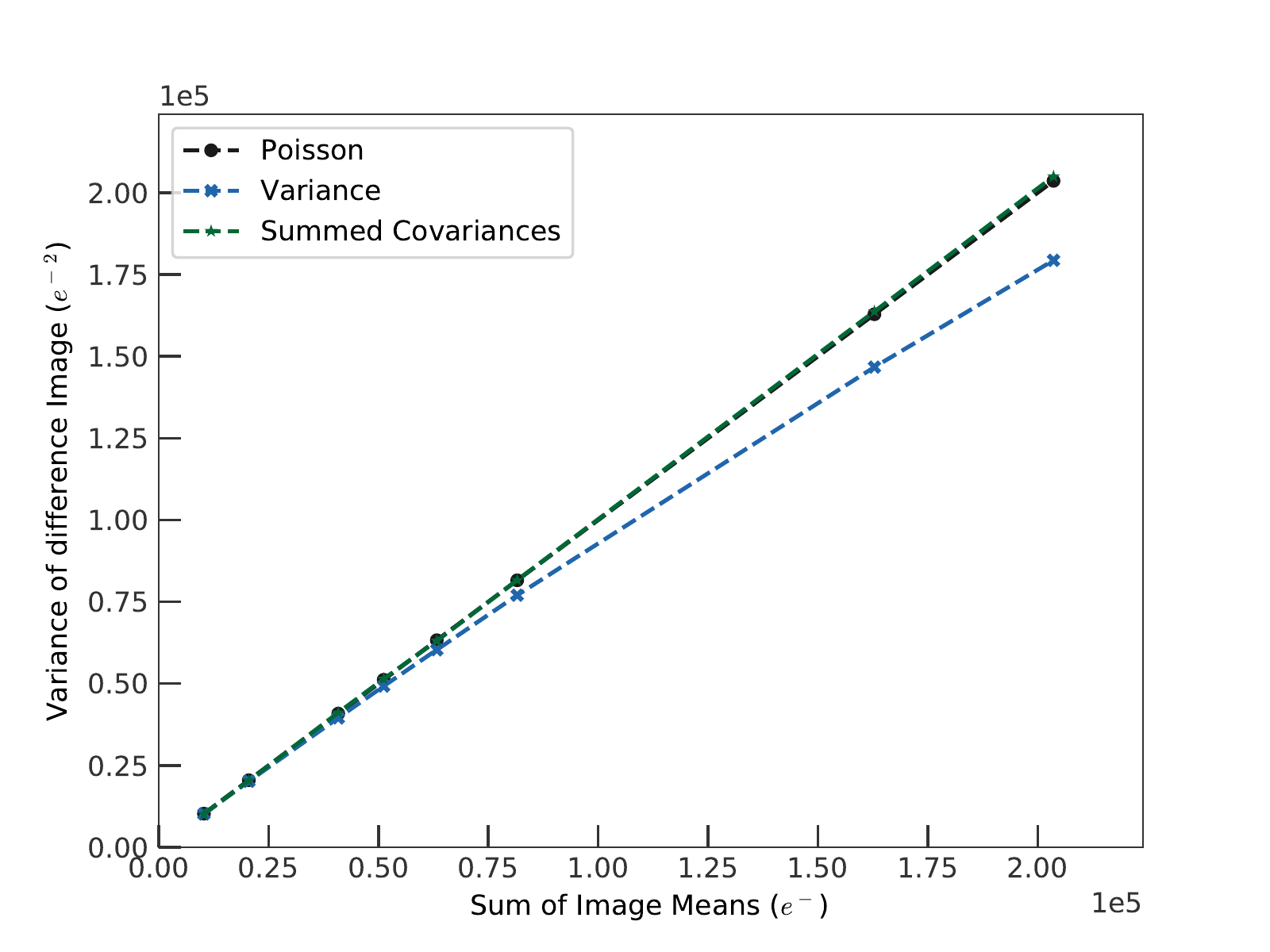}
	\caption{The photon transfer curve for CCD 40 obtained from analyzing difference images obtained from pairs of flat fields. It can be seen that the variance in the difference image deviates from the expected linear relation for images with long exposures (high means). The expected linear relation is regained if the pixel covariances are summed implying that the 'lost variance' has been transferred to the pixel-pixel correlations.}
	\label{fig:PTC_uncorr}
\end{figure}
\begin{figure}
\subfloat[Correlations in a 15 second exposure]{
  \centering
  \includegraphics[width=.5\textwidth]{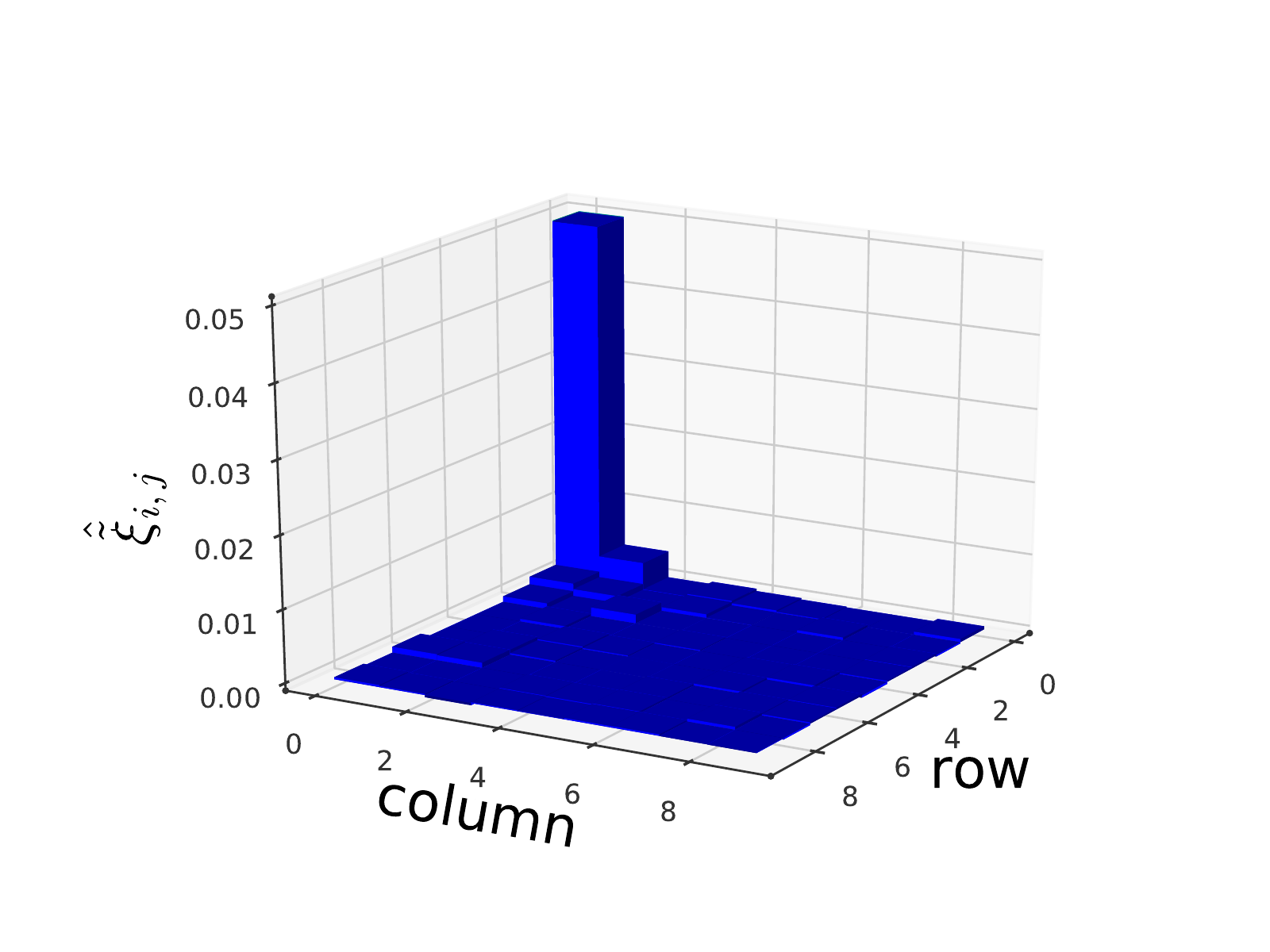}
} \qquad
\subfloat[Correlations in a 60 second exposure]{
  \centering
  \includegraphics[width=.50\textwidth]{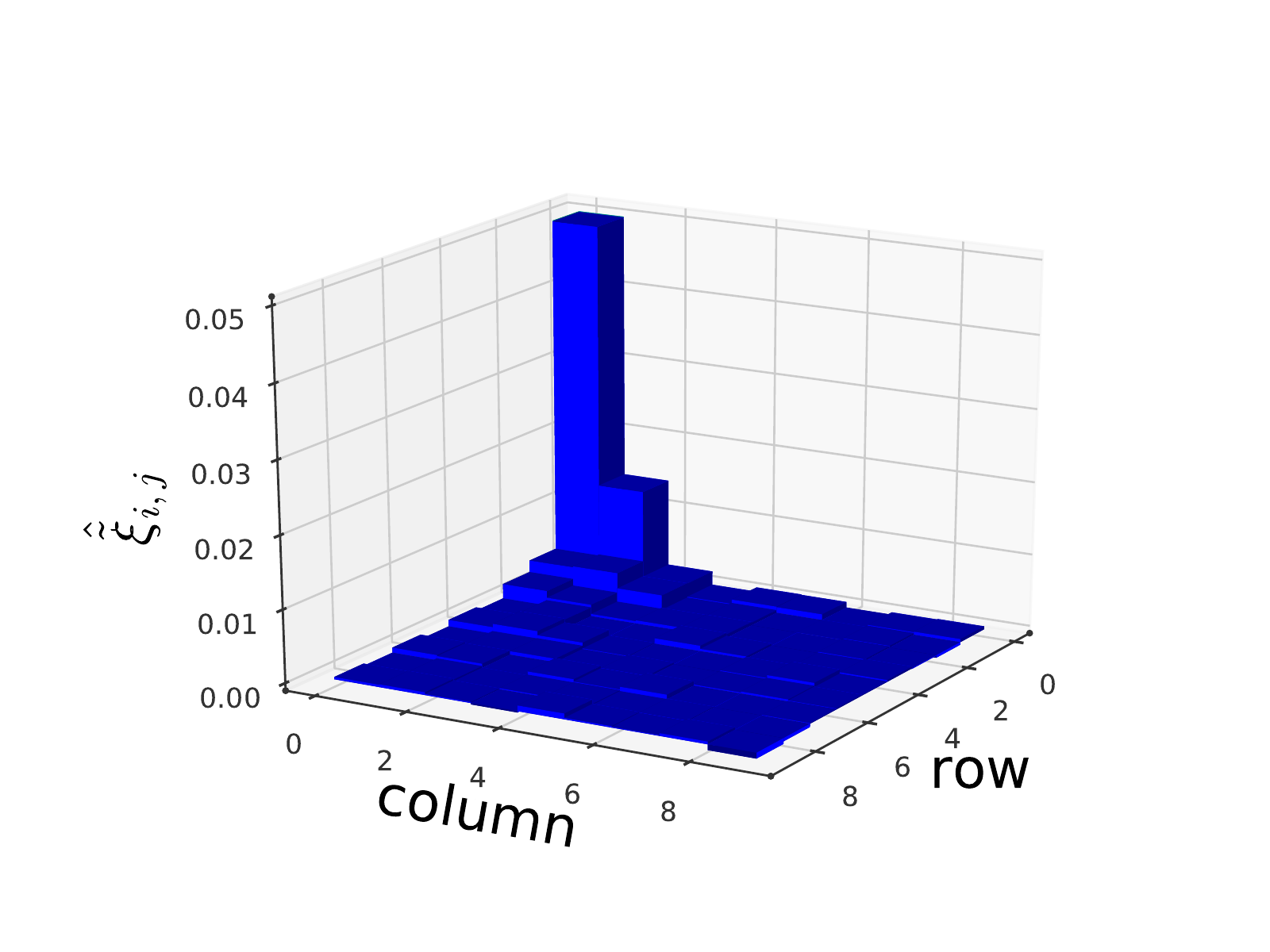}
}
\caption{Flat field correlation functions for images of two different durations, 15 and 60 seconds. The correlation functions are measured in the difference of two flat field images that have the same exposure times and are normalized by the mean of the image $\bar{F}$ as detailed in eq. \ref{eq:flatfieldcorrelation}. There exist correlations between the pixels and the strength of the correlations increase with exposure time.}
\label{fig:xcorr_uncorr}
\end{figure}

A photon transfer curve (PTC) shows how the variance in the signal changes with the mean flux level of a uniformly illuminated image, a flat field. In practice flat fields feature numerous instrumental effects that need to be removed before correlation functions can be measured. Here we briefly summarize how we process the flat field images. Firstly saturated pixels and defects are masked, we then apply crosstalk and non-linearity corrections before dark and bias frames are subtracted; the bias image corrects for the difference in each pixel's zero level while the dark image corrects for any thermally generated noise. Crosstalk is when signals from one amplifier induce correlations in adjacent amplifiers and is corrected via applying a correction matrix, which is the same for all the CCDs and was determined from commissioning data \citep{HSCpipeline}. After this the flat field images will still have the effects of variations in quantum efficiencies, non-uniform illumination (from both the variation in the source and from the optics) and other fixed effects. To reduce the impact of these in the flat fields, we difference of two images and then subtract a low spatial frequency background. From these difference images we calculate the variance and covariance of the pixels. 

In the ideal case the PTC should be a linear relation, once the read noise is negligible (HSC read noise is $\sim 4.5 e^{-}$ \citep{Nakaya2010}), with a slope of 1/gain until the pixels saturate. The linear relation arises due to the Poisson statistics of the incident photons. The loss of variance as the signal increases was first seen in \citep{downing2006}. HSC flat fields were taken for a 
range of different exposure times between 3 and  60 seconds and from these we attain PTCs; the PTC for  CCD 40 is  shown in figure \ref{fig:PTC_uncorr}. The flattening of the PTC (or the loss of variance) can be seen as can the dependence on integrated intensity.
We define the mean normalized correlation function as:
\begin{equation}\label{eq:flatfieldcorrelation}
\tilde{\xi}_{i,j}=\frac{ \langle (F_{I,J}-\bar{F}_{I,J} )(F_{I+i,J+j}-\bar{F}_{I+i,J+j} ) \rangle}{\bar{F}_{I,J}\bar{F}_{I+i,J+j}},
\end{equation} 
where $F_{I,J}$ is the counts recorded in the pixel (I,J). We calculate this by assuming that $\bar{F}_{I,J}=\bar{F}_{I+i,J+j}=\hat{\bar{F}}$, i.e. expected counts in a flat field are uniform across the amplifier and so is equal to the mean number of counts in the amplifier ($\hat{\bar{F}}=1/N_{\rm{pix}}\sum\limits_{I,J} F_{I,J}$). We calculate the correlation function as:
\begin{equation}
\hat{\tilde{\xi}}_{i,j}=\frac{1}{\hat{\bar{F}}^2 (N_{\rm{pix}}-1)} \sum\limits_{I,J} (F_{I,J}-\hat{\bar{F}} )(F_{I+i,J+j}-\hat{\bar{F}} ).
\end{equation} 
In figure \ref{fig:xcorr_uncorr} we plot the normalized correlation function for CCD 40 with exposures of duration 15 seconds and 60 seconds (i.e. mean counts of 25700 and 103000 e$^-$). It can clearly be seen that neighboring pixels are correlated in a slightly asymmetric way and that this effect is stronger for the longer exposure. Further the variance of these images is reduced below the expect value of the mean of the image, $\bar{F}$, to $0.961 \bar{F}$ and $0.899 \bar{F}$ for the 15 second and 60 second exposures.

\citet{Astier2014} noted that if the brighter fatter effect is responsible then the lost variance is transferred to correlations between pixels. The consequence of this is that the covariances can be summed together and this accounts for the lost variance (a similar result is shown in \citet{downing2006} where they considered the variance obtain after summing neighboring pixels into bigger pixels). In figure \ref{fig:PTC_uncorr} it can be seen that once summed together the expected linear relation is attained. One important consequence of this is that the gain can be obtained easily from just the gradient; once the covariances have been summed the PTCs are straight lines whose gradients are the inverse gains. 

\section{Theoretical Modeling}
 \label{ch:correction}

\citet{Astier2014} and \citet{Guyonnet2015} described an approach to correct this effect by modeling the effective pixel boundaries. Pixels with more charge than their neighbors have shifted effective boundaries as charges from photons incident near the boundary are deflected into adjacent pixels. They model the shift of the boundaries of the pixel at (0,0) due to the charge, Q$_{i,j}$, in the pixel at (i,j) as:
\begin{equation}
\delta^{X}_{i,j}=Q_{i,j}a^{X}_{i,j}
\end{equation}
where $\delta^{X}_{i,j}$ is the shift of the boundary,  $a^{X}_{i,j}$ characterizes how the boundary shifts due to the charge and the X superscript runs over the four edges of the pixel. The authors then assume that the $a^{X}_{i,j}$ are translationally invariant so the coefficients are the same for all pixels separated by (i,j). They use this formalism to model the brighter fatter effect and discuss how the effect could be corrected by measuring the $a^{X}_{i,j}$. To characterize the shifts caused by charge in pixels separated in x or y by less than n they need (n+1)(2n+4) coefficients. Flat fields correlation functions offer (n+1)$^2$ so this model cannot be constrained from flat fields without additional modeling. With the hope of circumventing this and for general broadness, we implemented an alternative method that will be discussed below. 
 \subsection{Camera Model}
 We modeled the camera in the limit of small pixels and it was treated as ideal photon counter. The measured positions on the focal plane of the $\mathrm{i}^{th}$  photon  are denoted by $\mathbf{x}_i$ and the time of arrival by $t_i$, where $t=0$ is the beginning of an exposure. The observed position on the focal plan has been deflected from the incident position, $\mathbf{y}_i$, by a deflection field, $\mathbf{\epsilon}(\mathbf{x}_i,t)$, such that $\mathbf{x}_i$=$\mathbf{y}_i$+$\mathbf{\epsilon}(\mathbf{x}_i,t)$. This deflection field, $\mathbf{\epsilon}$, is initially zero and grows as charge is accumulated. This field represents the idea that charge already accumulated in a pixel deflects other charge away from that pixel to its neighbors. 
 
 The 3D electric field $\mathbf{E}$ generated by the accumulated charge should have curl zero since the field is effectively static and this will lead to a curl free 2D deflection $\epsilon$. Given this, and also motivated by the parity inversion symmetry seen in the effect, it is assumed that the deflection field is curl free, $ \mathbf{\nabla_{x_i}} \times \mathbf{\epsilon}=0$. This means that $\mathbf{\epsilon}$ can be expressed as $ \mathbf{\epsilon}\propto  \mathbf{\nabla_{x_i}} K$. We also assume that this kernel, K, is translationally invariant and that the deflection field is proportional to the accumulated charge. Combining all these assumptions gives the following for the deflection field:
\begin{equation}
\mathbf{\epsilon}(\mathbf{x},t) = \sum_{t_i <t} \frac{\partial}{\partial x_i} K(\mathbf{x}-\mathbf{x_i}).
\end{equation}
This sum adds the contribution from all charges with arrival times, $t_i$, before $t$. Through the assumption that we can model this effect through a scalar potential we reduce the number of degrees of freedom to match the flat fields allowing our model to be constrained by flat field correlation functions alone.
\subsection{Consequences of the deflection field}
In the following sections we denote the incident intensity by $I(\mathbf{y})$ (assuming it to be time independent) and we model the photon events as a Poisson process with rate $I(\mathbf{y})$. We will d. We will be most interested in the properties of the counts field, which is defined as follows:
\begin{eqnarray}
\mathrm{F}(\mathbf{x},t)=\sum_{t_i <t} \delta^{(2)}(\mathbf{x}-\mathbf{x_i}) .
\end{eqnarray}
We define $\xi(\mathbf{x},\mathbf{x'},t)$, the correlation function between points $\mathbf{x}$  and $\mathbf{x'}$, as follows
\begin{eqnarray}
\xi(\mathbf{x},\mathbf{x'},t)= \langle \mathrm{F}(\mathbf{x},t) \mathrm{F}(\mathbf{x'},t) \rangle - \langle \mathrm{F}(\mathbf{x},t) \rangle \langle \mathrm{F}(\mathbf{x'},t) \rangle.
\end{eqnarray}
 We note that the measured Poisson process rate, $\lambda(\mathbf{x},t)$ will differ from the source, due to the deflection field, in the following way :
\begin{align}
\lambda(\mathbf{x,t})\mathrm{d}^2 \mathbf{x} = &I(\mathbf{y})\mathrm{d}^2 \mathbf{y},\\
\lambda(\mathbf{x,t})\mathrm{d}^2 \mathbf{x} =&( I(\mathbf{x}) - \epsilon_i(\mathbf{x})\frac{\partial I ( \mathbf{x})}{\partial x_i}) (1 -\frac{\partial \epsilon_i(\mathbf{x}))}{\partial x_i})\mathrm{d}^2 \mathbf{x},\\
\lambda(\mathbf{x,t})= &I(\mathbf{x})- \frac{\partial}{\partial x_i}\left(I(\mathbf{x})\frac{\partial}{\partial x_i} \int \mathrm{d}^2\mathbf{x'}K(\mathbf{x}-\mathbf{x'}) F(x',t)\right) , 
\label{eq:lambdadef}
\end{align}
where we enforce charge conservation in the first line and in the second line Taylor expand $I(\vec{x}-\epsilon(\vec{x}))$ for small deflections and use the Jacobian for the area element.  Utilizing the Poisson statistics of our events we see to first order that:
\begin{align}
\langle \mathrm{dF}(\mathbf{x}) \rangle =& \langle \mathrm{F}(\mathbf{x},t+dt) - \mathrm{F}(\mathbf{x},t) \rangle =\lambda(x,t) dt, \label{eq:dF}\\ 
\langle \mathrm{dF}(\mathbf{x})  \mathrm{dF}(\mathbf{x'}) \rangle  = &\lambda(x,t) \delta^{(2)}(\mathbf{x'}-\mathbf{x}) dt. \label{eq:dfdf}
\end{align}
Using the previous statements we can write down the time evolution equations for the mean : 
\begin{align}
\langle \mathrm{F}(\mathbf{x},t+dt) \rangle &= \langle \mathrm{F}(\mathbf{x},t) +\lambda(\mathbf{x},t) dt \rangle \\
&= \langle \mathrm{F}(\mathbf{x},t) \rangle + I(\mathbf{x})dt -\frac{\partial}{\partial x_i}\left(I(\mathbf{x})\frac{\partial}{\partial x_i} \int \mathrm{d}^2\mathbf{x'}K(\mathbf{x}-\mathbf{x'}) F(x',t)\right)dt 
\end{align}
and for the correlation function:
\begin{align}
\xi(\mathbf{x},\mathbf{x'},t+dt)=\langle  \mathrm{F}(\mathbf{x},t)  \mathrm{F}(\mathbf{x'},t)  + \mathrm{F}(\mathbf{x},t) d\mathrm{F}(\mathbf{x'},t)+ \mathrm{F}(\mathbf{x'},t) d\mathrm{F}(\mathbf{x},t) +d\mathrm{F}(\mathbf{x},t)d\mathrm{F}(\mathbf{x'},t)\rangle \nonumber \\ -\langle \mathrm{F}(\mathbf{x},t)+ d\mathrm{F}(\mathbf{x},t)\rangle \langle \mathrm{F}(\mathbf{x'},t)+d\mathrm{F}(\mathbf{x'},t)\rangle.
\end{align}
Note that the expectations on $  \mathrm{dF}(\mathbf{x}) $ are averages on events in $(t,t+dt)$ conditioned on the events in $(0,t)$ and so may be done first yielding.
\begin{align}
\xi(\mathbf{x},\mathbf{x'},t+dt)=\langle  \mathrm{F}(\mathbf{x},t)  \mathrm{F}(\mathbf{x'},t)  + \mathrm{F}(\mathbf{x},t)  \lambda(\mathbf{x'},t) dt +\mathrm{F}(\mathbf{x'},t)  \lambda(\mathbf{x},t) dt+ \lambda(\mathbf{x},t) dt \delta^{(2)}(\mathbf{x}-\mathbf{x'}) \rangle \\ -\langle \mathrm{F}(\mathbf{x'},t)+  \lambda(\mathbf{x'},t) dt\rangle \langle \mathrm{F}(\mathbf{x},t) + \lambda(\mathbf{x},t) dt)\rangle \\
\xi(\mathbf{x},\mathbf{x'},t+dt)=\xi(\mathbf{x},\mathbf{x'},t)- \frac{\partial}{\partial x_i}\left(I\left(\mathbf{x}\right)\frac{\partial}{\partial x_i} \int \mathrm{d}^2\mathbf{x''}K\left(\mathbf{x}-\mathbf{x''}\right) \xi\left(\mathbf{x''},\mathbf{x'},t\right)\right)dt \nonumber \\ - \frac{\partial}{\partial x'_i}\left(I\left(\mathbf{x'}\right)\frac{\partial}{\partial x'_i} \int \mathrm{d}^2\mathbf{x''}K\left(\mathbf{x'}-\mathbf{x''}\right) \xi\left(\mathbf{x},\mathbf{x''},t\right)\right)dt
 + \left(\langle \mathrm{F}(\mathbf{x},t+dt) \rangle - \langle \mathrm{F}(\mathbf{x},t) \rangle)\right)\delta^{(2)}\left(\mathbf{x}-\mathbf{x'}\right) dt.
\end{align}
From which we attain the following differential equations:
\begin{align}
\frac{\partial \langle \mathrm{F}\left(\mathbf{x},t\right) \rangle }{\partial t} = I\left(\mathbf{x}\right) - \frac{\partial}{\partial x_i}\left(I\left(\mathbf{x}\right)\frac{\partial}{\partial x_i} \int \mathrm{d}^2\mathbf{x'}K\left(\mathbf{x}-\mathbf{x'}\right) \langle \mathrm{F}\left(\mathbf{x'},t \right) \rangle\right) 
\end{align}
\begin{align}
\frac{\partial \xi\left(\mathbf{x},\mathbf{x'},t\right) }{\partial t}= - \frac{\partial}{\partial x_i}\left(I\left(\mathbf{x}\right)\frac{\partial}{\partial x_i} \int \mathrm{d}^2\mathbf{x''}K\left(\mathbf{x}-\mathbf{x''}\right) \xi\left(\mathbf{x''},\mathbf{x'},t\right)\right) - \frac{\partial}{\partial x'_i}\left(I\left(\mathbf{x'}\right)\frac{\partial}{\partial x'_i} \int \mathrm{d}^2\mathbf{x''}K\left(\mathbf{x'}-\mathbf{x''}\right) \xi\left(\mathbf{x},\mathbf{x''},t\right)\right)\nonumber
\\ 
+\left(I\left(\mathbf{x}\right) - \frac{\partial}{\partial x_i}\left(I\left(\mathbf{x}\right)\frac{\partial}{\partial x_i} \int \mathrm{d}^2\mathbf{x''}K\left(\mathbf{x}-\mathbf{x''}\right) \langle \mathrm{F}\left(\mathbf{x''},t\right) \rangle \right)\right)\delta^{(2)}\left(\mathbf{x}-\mathbf{x'}\right).
\end{align}
Under the assumption that K is small we can solve these equations iteratively, giving to first order:

\begin{equation}
\langle \mathrm{F}\left(\mathbf{x},t\right) \rangle = tI\left(\mathbf{x}\right) - \frac{t^2}{2} \frac{\partial}{\partial x_i}\left(I\left(\mathbf{x}\right) \frac{\partial}{\partial x_i} \int \mathrm{d}^2\mathbf{x'}K\left(\mathbf{x}-\mathbf{x'}\right) I\left(\mathbf{x'} \right) \right) \\
\label{eq:pointFunctions}
\end{equation}
\begin{equation}
\xi(\mathbf{x},\mathbf{x'},t)= \langle \mathrm{F}\left(\mathbf{x},t\right) \rangle\delta^{(2)}(\mathbf{x}-\mathbf{x'})-\frac{t^2}{2}  \frac{\partial}{\partial x_i}\left( I(\mathbf{x})  \frac{\partial}{\partial x_i}K(\mathbf{x}-\mathbf{x'})\right) I(\mathbf{x'})-\frac{t^2}{2}  \frac{\partial}{\partial x'_i}\left( I(\mathbf{x'})  \frac{\partial}{\partial x'_i} K(\mathbf{x'}-\mathbf{x})\right) I(\mathbf{x}).
\label{eq:pointFunctions2}
\end{equation}
These show the influence of the deflection field on the expected counts and the covariances. In particular we note that for the case of an ideal flat field ($I(\mathbf{x})=I_0$) the correlation function has the following form:
\begin{equation}
\xi(\mathbf{x},\mathbf{x'},t)= I_0\delta^{(2)}(\mathbf{x}-\mathbf{x'})-t^2 I_0^2  \frac{\partial^2}{\partial x^2_i}K(\mathbf{x}-\mathbf{x'}),
\label{eq:flatfieldcor}
\end{equation}
where we assumed parity symmetry to simplify the result. Thus if  $\left.  \partial^2 K(x-x') / {\partial x^2_i}\right|_{x=x'}>0$ the variance will be reduced. Assuming that $K$ goes to zero at large separations we have:
\begin{equation}
\int d^2 x'   \frac{\partial^2}{\partial x^2_i }K(\mathbf{x}-\mathbf{x'}) = 0.
\end{equation}
This sanity-check confirms that charge will be conserved and so the mean of the flat field is unchanged by the brighter fatter effect. 
\subsection{The proposed correction}

After considering the form the deviation shown in eq. \ref{eq:pointFunctions}, a logical proposal for recovering the incident image is to consider the following estimator:
\begin{equation}
\hat{\mathrm{F}}(\mathbf{x},t) =  \mathrm{F}\left(\mathbf{x},t\right)+ \frac{1}{2} \frac{\partial}{\partial x_i}\left( \mathrm{F}\left(\mathbf{x},t\right)  \frac{\partial}{\partial x_i} \int \mathrm{d}^2\mathbf{x'}K\left(\mathbf{x}-\mathbf{x'}\right) \mathrm{F}\left(\mathbf{x'},t\right)\right)
\label{eq:correction}
\end{equation}
It is immediately seen, after we substitute from eq. \ref{eq:pointFunctions} and only keep the first order terms in K, that:
\begin{equation}
\langle \hat{\mathrm{F}}(\mathbf{x},t)\rangle= t I(\mathbf{x}).
\end{equation}
Thus the correction restores the mean to the expected value. Now considering the covariance  of our estimator (to first order in K):
\begin{eqnarray}
\mathrm{Cov}(\hat{\mathrm{F}}(\mathbf{x},t),\hat{\mathrm{F}}(\mathbf{x'},t) )= \xi(\mathbf{x},\mathbf{x'},t) +\frac{1}{2} \frac{\partial}{\partial x_i}\left(\int \mathrm{d}^2\mathbf{x''} \mathrm{Cov}\left(\mathrm{F}(\mathbf{x})\mathrm{F}(\mathbf{x''}),\mathrm{F}(\mathbf{x'})\right) \frac{\partial}{\partial x_i}K(\mathbf{x}-\mathbf{x''})\right) \nonumber\\
+\frac{1}{2}\frac{\partial}{\partial x'_i}\left(\int \mathrm{d}^2\mathbf{x''} \mathrm{Cov}\left(\mathrm{F}(\mathbf{x'})\mathrm{F}(\mathbf{x''}),\mathrm{F}(\mathbf{x})\right) \frac{\partial}{\partial x'_i}K(\mathbf{x'}-\mathbf{x''})\right).
\end{eqnarray}
Again substituting from eqs. \ref{eq:pointFunctions} and \ref{eq:pointFunctions2}, keeping only first order terms and using $\frac{\partial}{\partial x_i} K(0)=0$ 
(that there is no deflection at the exact point of the charge build up, a condition that is also forced by the parity symmetry) we arrive at the following:
\begin{equation}
\mathrm{Cov}(\hat{\mathrm{F}}(\mathbf{x},t),\hat{\mathrm{F}}(\mathbf{x'},t) )= t I(\mathbf{x})\delta^{(2)}(\mathbf{x}-\mathbf{x'}).
\end{equation}
This shows that the correction proposed in eq. \ref{eq:correction} undoes the charge redistribution of the brighter fatter effect to first order in the deflection kernel.  To use this correction the kernel needs to be known and it can be attained from flat fields using eq. \ref{eq:flatfieldcor}.

It should be noted that in our applications of the correction we used the following form:
\begin{equation}
\frac{1}{2} \frac{\partial}{\partial x_i}\left( F(\mathbf{x})\right)\frac{\partial}{\partial x_i} \int \mathrm{d}^2\mathbf{x'} F(\mathbf{x'})K(\mathbf{x}-\mathbf{x'}) + \frac{1}{2} F(\mathbf{x}) \frac{\partial^2}{\partial x^2_i}  \int \mathrm{d}^2\mathbf{x'} F(\mathbf{x'})K(\mathbf{x}-\mathbf{x'})
\label{eq:correction2}
\end{equation}
and, for our pixelated data, we implement the first order derivatives using finite central differences and the second order derivatives using two successive forward finite differences. The higher accuracy obtained using the two successive forward finite differences motivates the use of eq. \ref{eq:correction2}. 

To first order the distortion effect depends upon the 'incident counts'  (those that would be recorded in an ideal detector) and our first order correction uses the measured counts (eq. \ref{eq:correction} ). By using the measured counts a higher order term is induced and that would be removed if the 'incident counts' were used. To this end we use an iterative procedure that reconstructs the incident counts and uses them in the correction. To do this we calculate eq. \ref{eq:correction2} and add this to copy of the image. We calculate the correction from this modified image and add it to a copy of the original image. This is repeated until the correction from the modified image converges. Through this iterative process we obtain increasingly accurate estimates of the 'incident counts' and the difference between our correction and the first order distortion decreases.

\section{Application to flat fields}
\label{ch:flatfield}
\subsection{Recovering the kernel}
Eq. \ref{eq:flatfieldcor} implies that from the correlation function of an ideal flat field we should be able to simply recover the kernel. As described in section \ref{ch:HSCPTC} we use difference flat field images to measure covariances. Ignoring any effects due to the image processing, we extend the results of the previous section to attain an expression for the kernel from these differenced flat field correlation functions. In particular, we apply eq. \ref{eq:pointFunctions2} to the case where $I(\mathbf{x})=I_0$ and we substitute this into the correlation function:
\begin{eqnarray}
 \mathrm{Cov}(\mathrm{F}_A(\mathbf\mathbf{x})) -\mathrm{F}_B(\mathbf{x}) , \mathrm{F}_A(\mathbf{x'}) -\mathrm{F}_B(\mathbf{x'})) = \mathrm{Cov}(\mathrm{F}_A(\mathbf{x}) ,\mathrm{F}_A(\mathbf{x'})) + \mathrm{Cov}(\mathrm{F}_B(\mathbf{x}) ,\mathrm{F}_B(\mathbf{x'})) \nonumber \\
 = tI_A \delta^{(2)}(\mathbf{x}-\mathbf{x'}) +t I_B \delta^{(2)}(\mathbf{x}-\mathbf{x'})-t^2({I_{A}}^2 + {I_{B}}^2) \frac{\partial}{\partial x_i} \cdot\frac{\partial}{\partial x_i}(K(\mathbf{x}-\mathbf{x'}))
\label{eq:correlation2}
\end{eqnarray}
where $\mathrm{F}(\mathbf{x})_A$ and $\mathrm{F}(\mathbf{x})_B$ refer to the counts in the two images used in the subtraction, which are assumed to be independent, and $I_A$ and $I_B$ are the intensities image A and B were exposed to for duration t. This equation can 
be numerically inverted using successive over relaxation (SOR) to attain the kernel, with the kernel assumed to go to zero on the boundary and 
$tI_A$ and $tI_B$ approximated by the mean counts in each image. 

We used flat field data obtained between April 2013 and December 2016 from all the broad bands filters. This data set consisted of flat field images with exposure times between 3 and 96 seconds, with many exposures with duration between 10 and 20 seconds.

\begin{figure}  
\subfloat[X cross section]{
  \centering
\includegraphics[width=.5\textwidth]{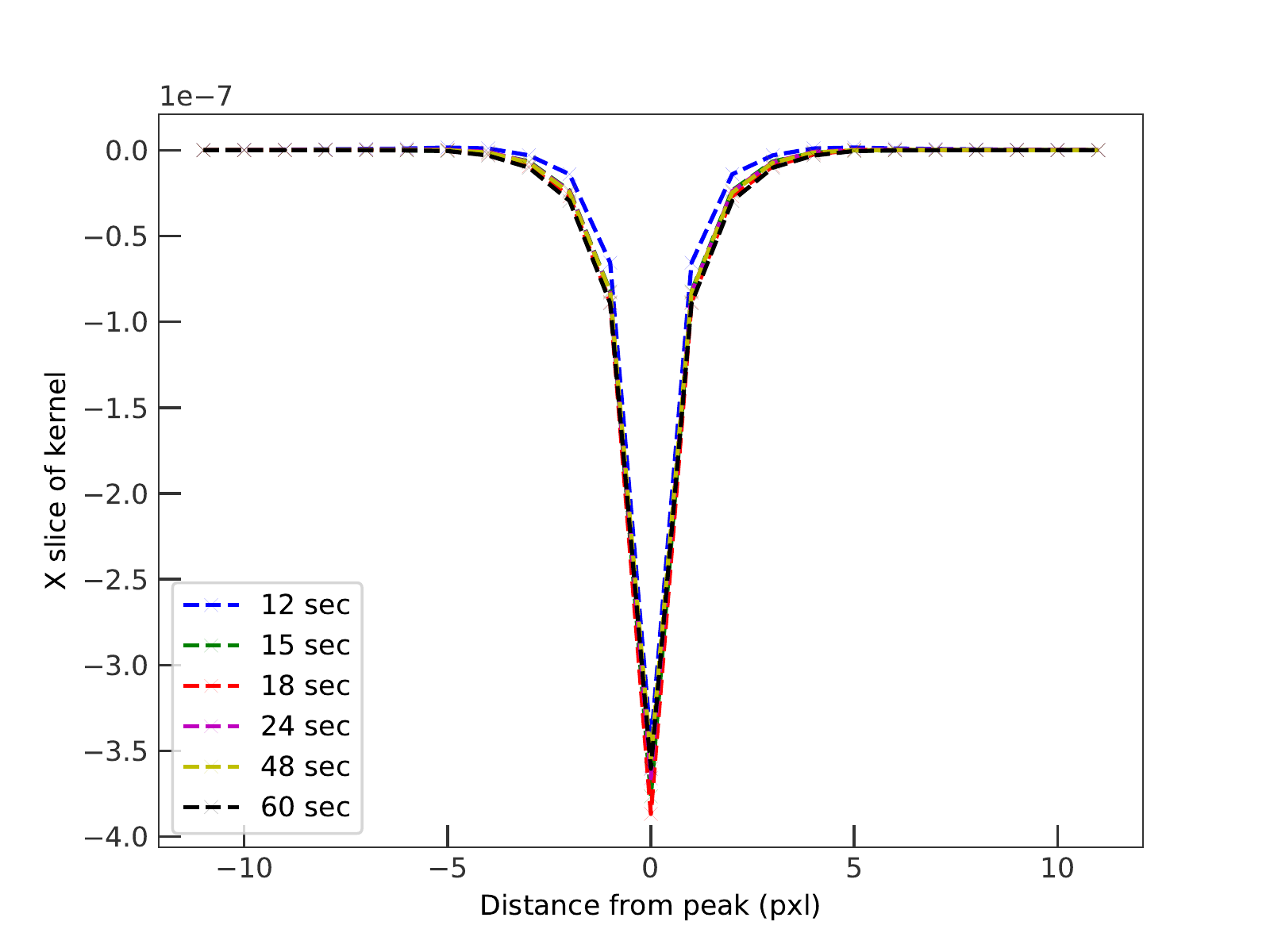}
\label{fig:kernel_crossX}
} \qquad
\subfloat[Y cross section]{
  \centering
\includegraphics[width=.5\textwidth]{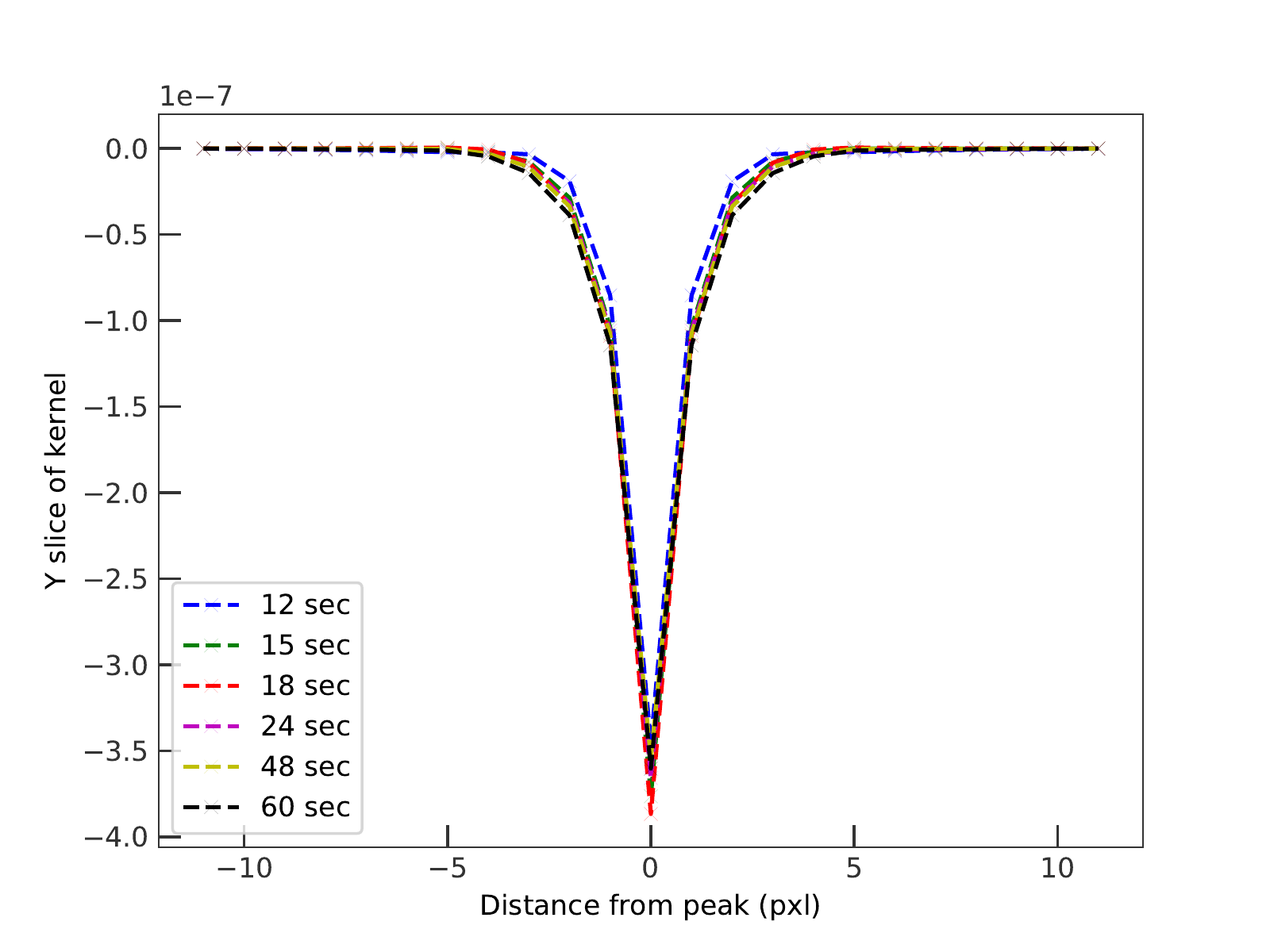}
\label{fig:kernel_crossY}
}
\caption{A cross section through the peak of the recovered kernel for a set of different exposure times. We expect that the recovered kernel should be independent of brightness as so the same kernel should be obtained from flat fields with different exposure times. Whilst there is some scatter, it can be seen that there is no significant correlation with exposure time. }
\label{fig:kernel_cross}
\end{figure}

Figure \ref{fig:kernel_cross} shows a slice through the kernels recovered from flat fields with different exposure times. These kernels are attained after averaging over all of the CCDs; this was done to attain a more robust measurement kernel measurement as theoretically the kernel should not vary significantly across the camera. The results show some scatter but there is no strong dependence on the intensity of the flat fields. This is an important self consistency check as theoretically the kernel should not depend on the intensity of the flat fields used to measure it. For the remainder of the analysis we will use a kernel that is attained using all of these exposure times.

\subsection{Flat field results}
\begin{figure}
\subfloat[Correlations in an uncorrected 60 second exposure.]{
  \centering
 \includegraphics[width=.50\textwidth]{./Xcorr_visit_904646_904648_ccd_40.pdf}
 \label{fig:xcorr_uncorr_comp}
}
\subfloat[Correlations in a corrected 60 second exposure.]{
  \centering
 \includegraphics[width=.50\textwidth]{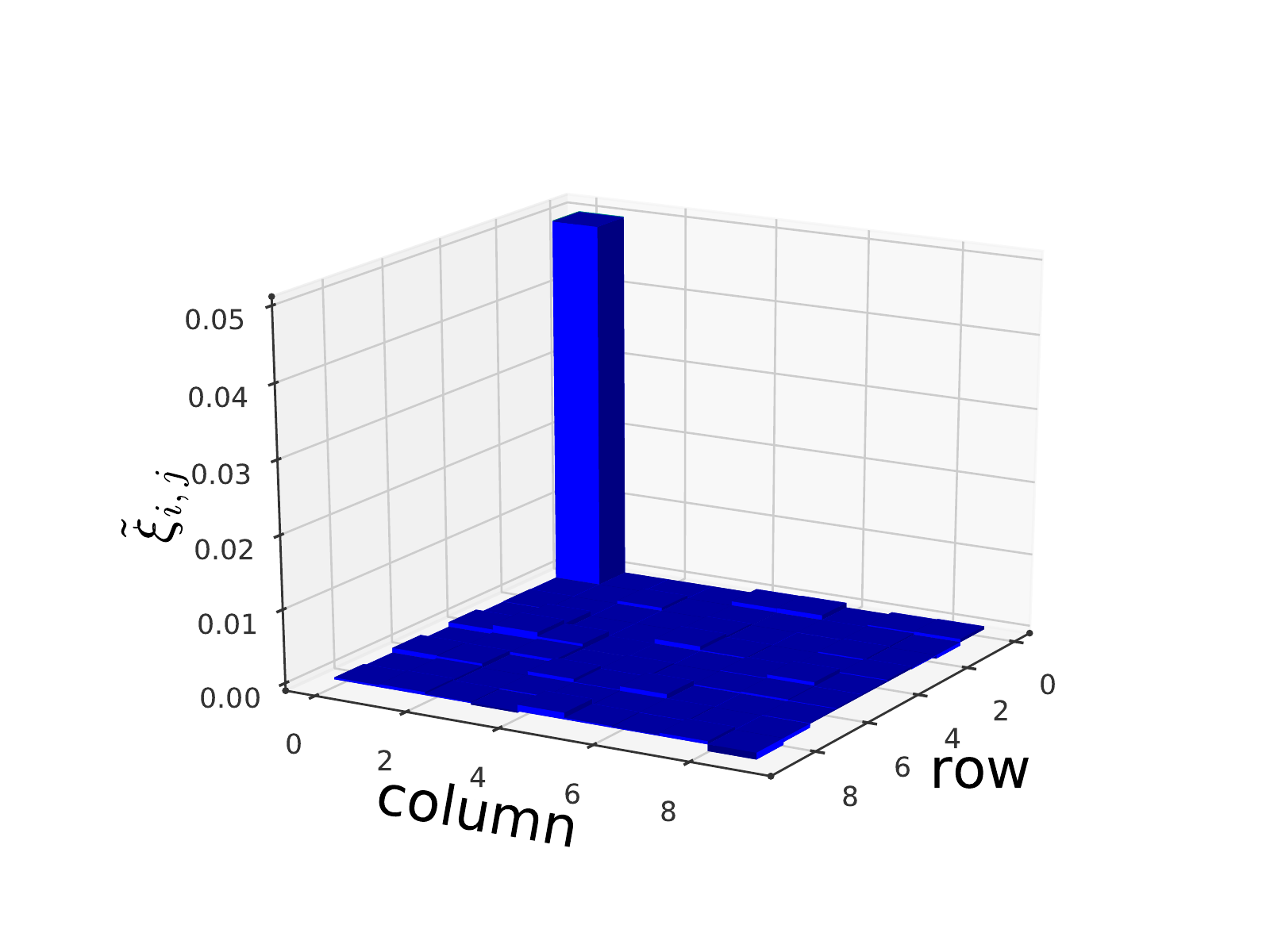}
\label{fig:xcorr_corr_bad}
}
\caption{ In figure (a) we plot the normalized correlation function in a pair of  flat field images with 60 second exposures. Figure (b) shows the correlation function of the same images after applying our brighter-fatter correction, eq. \ref{eq:correction2}. It can be seen that the correction removes almost all of the pixel correlations. }
\end{figure}

Figure \ref{fig:xcorr_uncorr_comp} again shows the normalised correlation function for a pair of flat fields with a 60 second exposure in CCD 40.  The asymmetry can be explained by the CCD structure, the pixel boundaries in the row direction are set by channel stops, implanted dopants, whereas they are set by gate voltages in the column direction and these differences make it likely that the transverse electric field in these directions will be different.  Figure \ref{fig:xcorr_corr_bad} shows the correlation function measured in the same images after we have applied our correction. It can be seen that almost all of the correlations between pixels have been removed. Further the variance of the image is now $1.007 \bar{F}$ (before the correction it was $0.899 \bar{F}$)  which agrees with the expectation for a Poisson relation, $1.0 \bar{F}$.

\begin{figure}
  \centering
  \includegraphics[width=.5\textwidth]{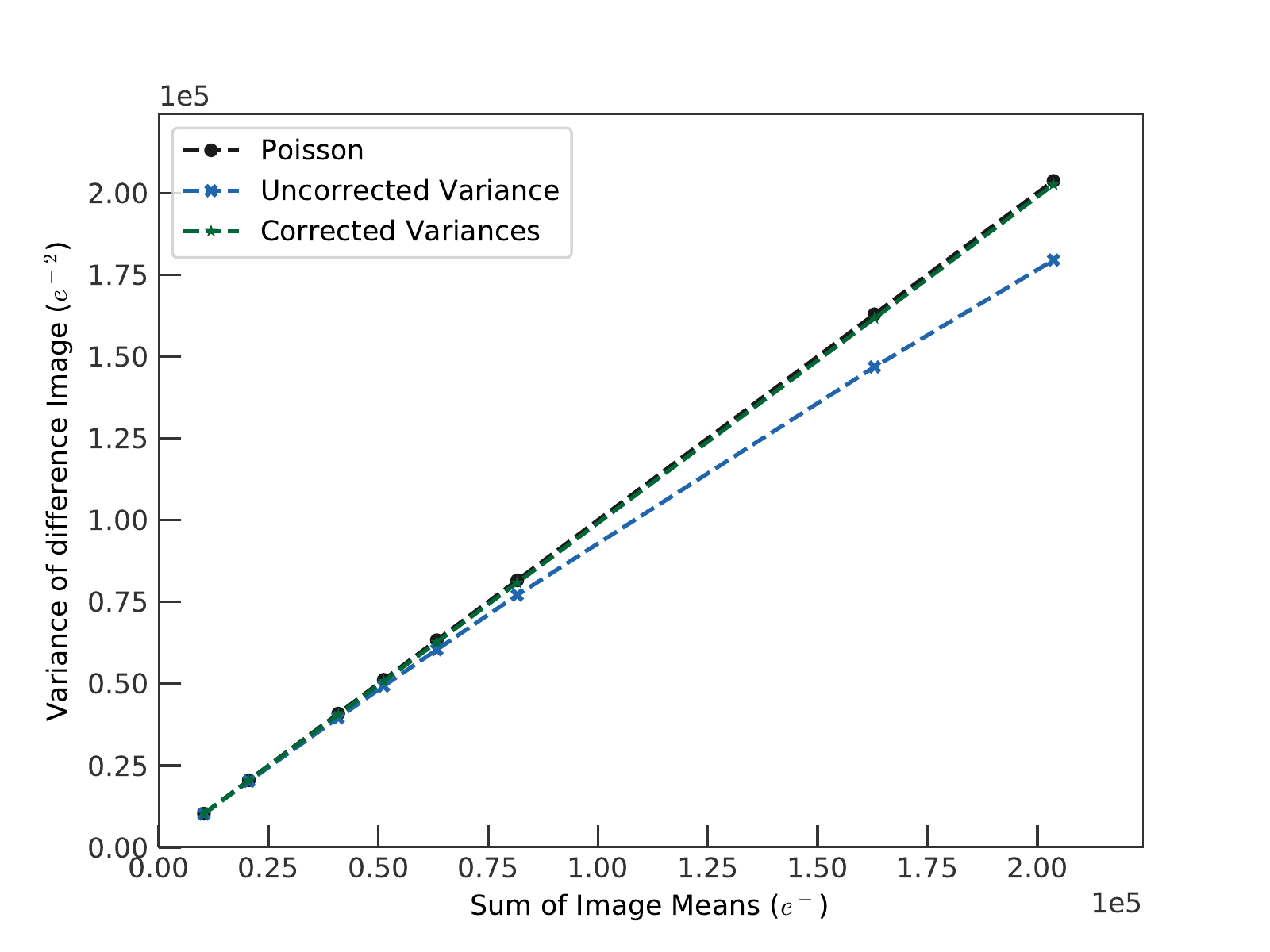}
   \caption{The photon transfer curve for CCD 40 obtained after processing the images with our brighter-fatter correction, eq. \ref{eq:correction2}. It can be seen that our correction restores the expected linear relation .}
 \label{fig:corrected_PTC}
\end{figure}

Having applied our correction to all of the flat fields we then regenerated the PTC and this is shown in figure \ref{fig:corrected_PTC}. It can seen that the correction restores the expected linear relation to a high level. Whilst this is a good result for our method, it is unsurprising. Examining the form of eqs. \ref{eq:correction2} and \ref{eq:flatfieldcor}  it can be seen that we are essentially deconvolving the flat field covariances. Numerically this is confirmed, as the first term in eq. \ref{eq:correction2} can be seen to be subdominant to the second term. 

\section{Application to stars}
\label{ch:stars}

\begin{figure}
\subfloat[Before correction]{
  \centering
  \includegraphics[width=.450\textwidth]{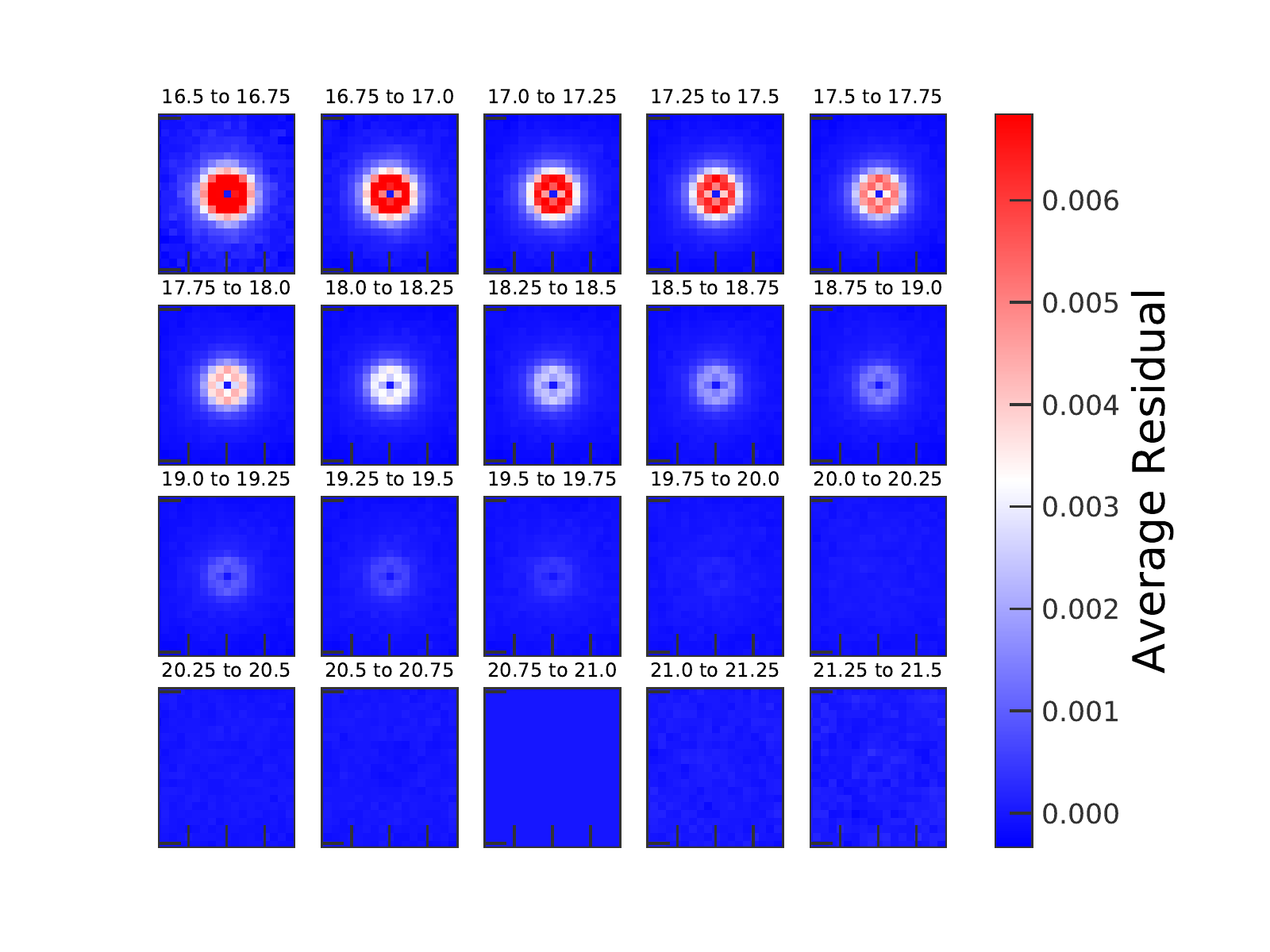}
\label{fig:2dResiduals}
}\qquad
\subfloat[After correction]{
  \centering
\includegraphics[width=.450\textwidth]{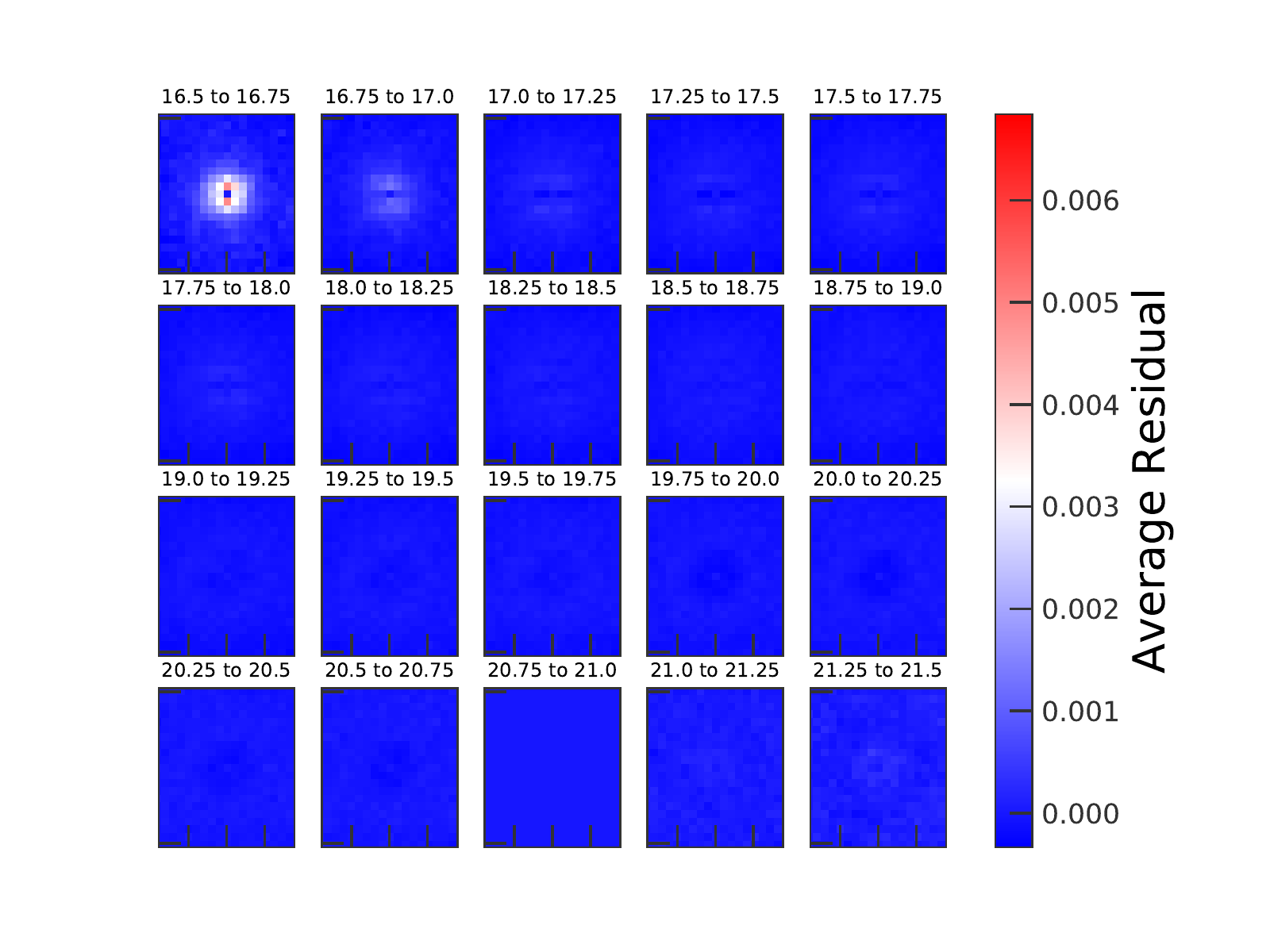}
\label{fig:2dResidualsAfterCorrection}
}
   \caption{The 2D residuals when the average star of the stars with $20.25<M_{\rm{instrm}}<20.5$ is subtracted from stars binned by instrumental magnitude and normalized by their peak value. The brightest objects are in the top left and the faintest objects in the bottom right. Without the correction the brightest stars are broadened when compared to the faintest stars and so we an excess residual. After our correction we see the residuals are significantly reduced for a large range of magnitudes.}
 \label{fig:starResiduals}
\end{figure}

\begin{figure}
\centering
\includegraphics[width=.50\textwidth]{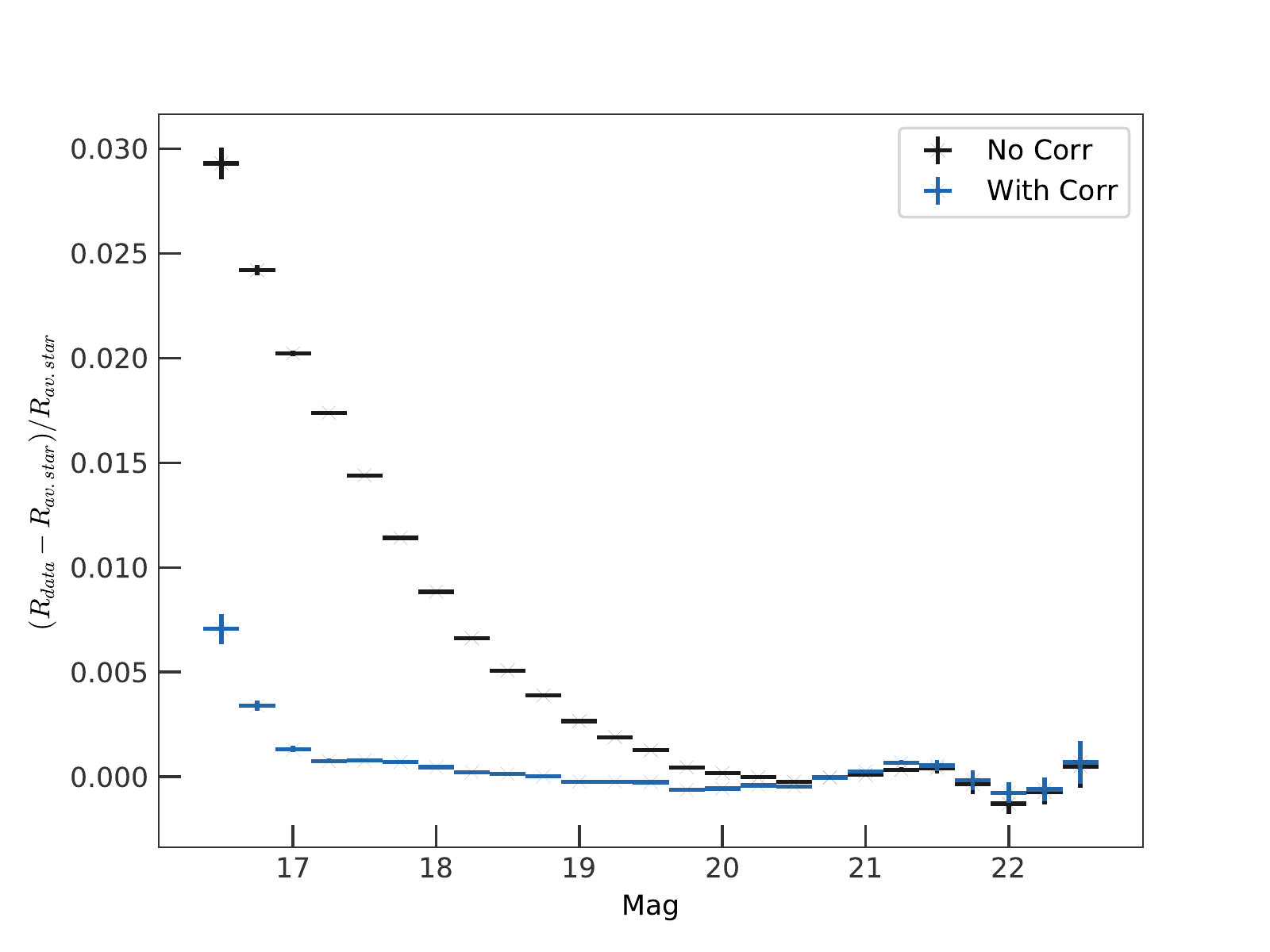}
 \caption{A graph showing how the radii residuals of stars, compared to the average star as described in the text, varies as a function of stellar brightness. The brighter fatter effect broadens the stars and so we see brighter stars having a larger radius. It can be seen that our correction removes most of the luminosity dependence of the effect.}

\label{fig:starRadii_noPSF}
\end{figure}
\begin{figure}
\subfloat[$e_1$ ellipticity]{
\centering
\includegraphics[width=.50\textwidth]{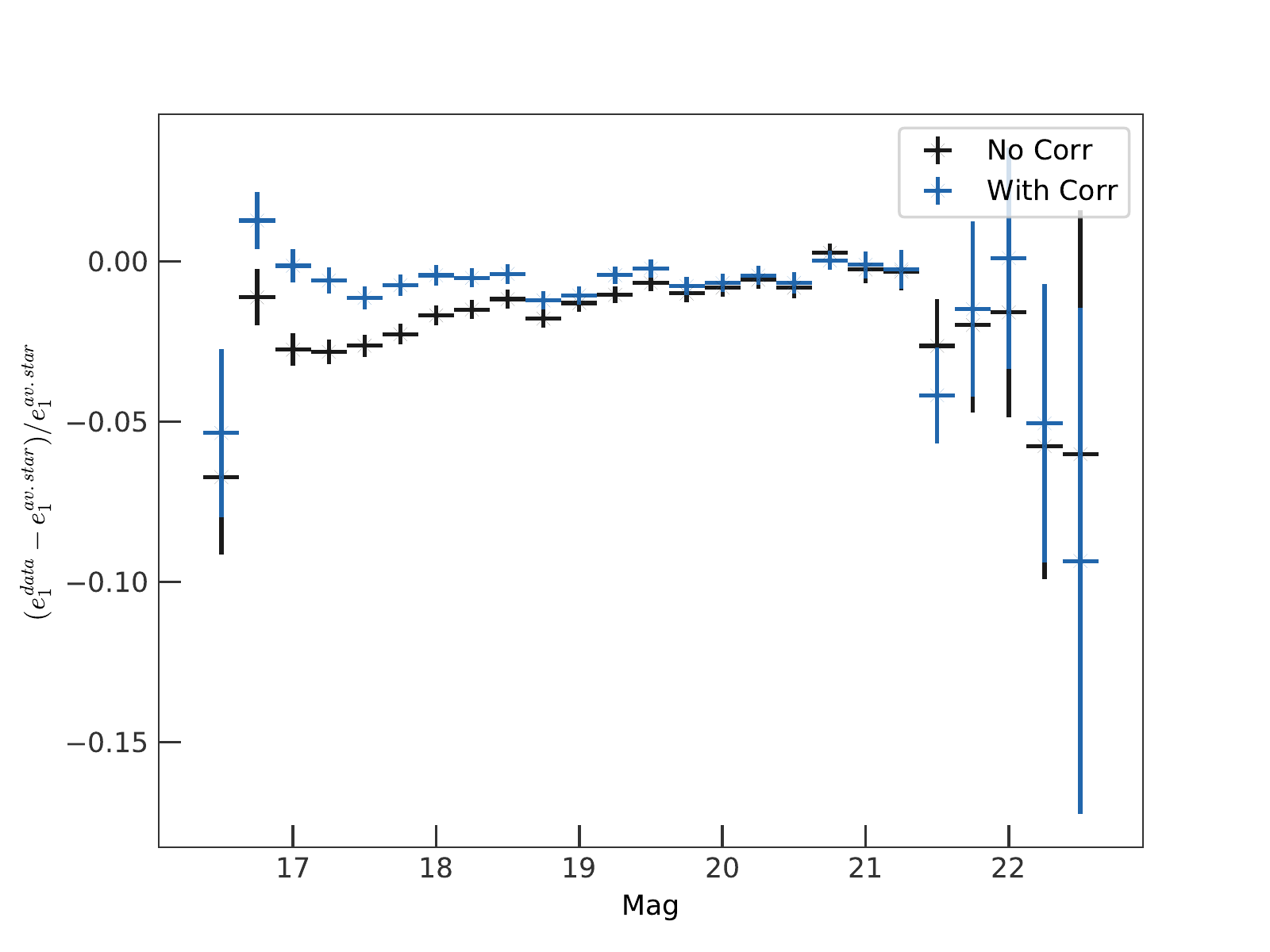}
\label{fig:e1Moments}
}\qquad
\subfloat[$e_2$ ellipticity]{
  \centering
\includegraphics[width=.50\textwidth]{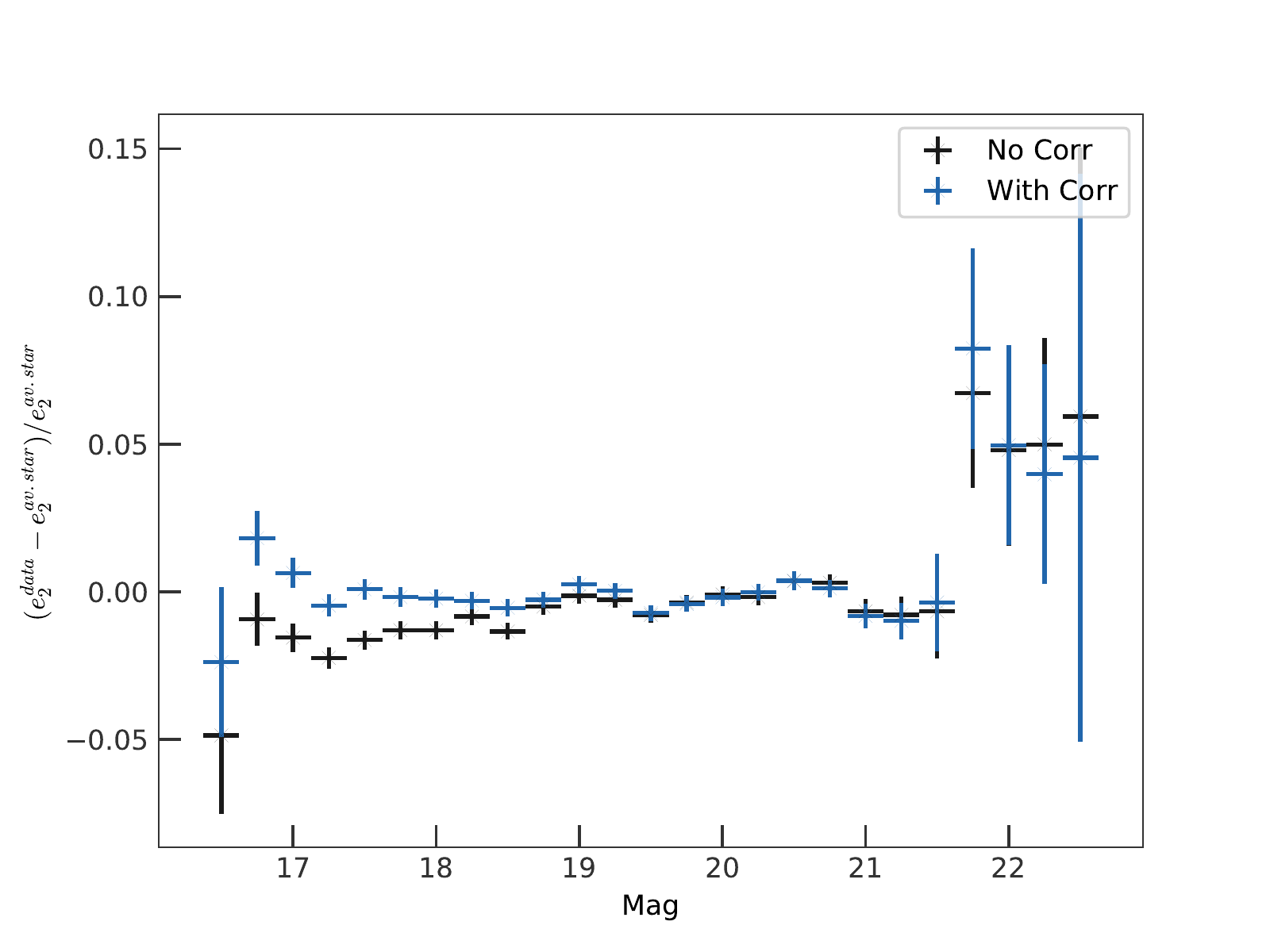}
\label{fig:e2Moments}
}
\caption{ The shape dependence of the stars with brightness is explored with the two ellipticity measures described in the text. We plot the fractional difference of the ellipticity between stars and the average star  both with and without the correction. The brighter fatter effect produces an ellipticity trend with luminosity as the CCD structure is different in the row and column direction, with the pixel boundaries having different physical sources. We find that our correction removes most of the luminosity trend. }
\label{fig:eMoments_noPSF}
\end{figure}
The second major consequence of brighter fatter effect is the intensity dependence of the PSF.  The simplest way to see this is, as is demonstrated in \citet{Lupton2014},
when a scaled dim star is subtracted from a bright star as there is a large residual pattern.  The residuals arise as the brighter fatter effect broadens the PSF and distorts the brighter stars' shapes from
the dimmer stars and this is shown in figure \ref{fig:2dResiduals}. Note that  in this work we use instrumental magnitudes $-2.5 log{F}+31.5$  as the brighter fatter effect is only sensitive to the number of holes measured and not to the zero point offset.  We quantified the distortion of the stars by examining how the second moments of the stars depend on their brightness. We calculate the adaptive gaussian second moments of the stars \citep{Bernstein2002} and the implementation is described in \citet{HSCpipeline}. In this method we first calculate:
\begin{equation}
M_{i,j}=\frac{\int \mathrm{d}^2\theta I(\vec{\theta})\theta_i\theta_j W(\vec{\theta})}{{\int \mathrm{d}^2\theta I(\vec{\theta})}},
\label{eq:adpMom}
\end{equation}
where $I(\vec{\theta})$ is the object intensity at position $\theta$, $W(\vec{\theta})$ is a gaussian weight used to ensure the integral converges, and $\theta_{i/j}$ is  $\theta_{x/y}$ depending on which moment $M_{xx},M_{xy},M_{yy}$ is being computed. The width of the gaussian is initially guessed, the moments are calculated, the moments are then used as the width in the weights' gaussian and the process is iterated until the weights and the moments are equal. In this work we use the following definitions for objects' radii and ellipticities:
\begin{align}\label{eq:radius}
R=\left(M_{xx}^2+M_{yy}^2\right)^{\frac{1}{4}} \\
e_1=\frac{M_{xx}-M_{yy}}{M_{xx}+M_{yy}} \\ 
e_2=\frac{2M_{xy}}{M_{xx}+M_{yy}} 
\end{align}

We first process the visits as described in \citet{HSCpipeline} and then use all the objects selected for use in the PSF model. We use a selection of visits taken between September 2014 and March 2016 from all the broad bands and from a range of seeings (discussed further below). The visits are a combination of 30, 150 and 200 second exposures from the wide survey, which will cover two long stripes on the equator and a stripe around the Hectomap region \citep{HSCsurveyOverview}.  In order to reduce the shape noise we need to average our measurements over many stars. The different conditions, and focal plane locations, of our measurements lead to variations in stellar radii and shapes and in order to disentangle these effects we use two approaches: in the first method we construct an average star for each visit and CCD, hereafter called the average star method, and compare our measurements against this star; in the second method we compare our measurements to PSF model, which encapsulates the previously mentioned effects. 

\subsection{Average Star Method}

In this method we select all the stars  with $20.25<M_{\rm{instrm}}<20.5$ and then we compute, for each exposure and each CCD, the average properties of this bin. We then calculate differences between this average star and stars in other luminosity bins. The differences (and fractional quantities) in this method are less sensitive to variations in conditions and telescope effects but the fractional quantities suffer from increased noise due to statistical noise in the average star.

We plot the fractional residuals of stars compared to the average star in figure \ref{fig:starRadii_noPSF}, for stars observed  in the \textit{i} band with $0.61"<$ FWHM $<=0.66"$. Before the correction we see that brighter stars have much larger radii than the fainter stars.  After applying our correction, it can be seen that only a small residual trend  remains for the brightest stars. The intrinsic stellar ellipticity can be altered by the brighter fatter effect if the kernel is asymmetric and/or the PSF is asymmetric, both of which are the case in the HSC data.  In figure \ref{fig:eMoments_noPSF} we plot the fractional ellipticity difference for the two ellipticities as a function of instrument magnitude. The shapes can be seen to be distorted as the brightness increases. In the same figure we show the difference after applying our correction; we see that the trend is completely removed though the errors are still moderately large and so more data is required to verify this. In figure \ref{fig:2dResidualsAfterCorrection} we show the 2D residuals after applying our correction. In this figure you can see the 2D residual between the average star in that bin and the average star with $22.25<M_{{instrm}}<22.5$ as a function of luminosity.

We see that our correction leaves a small residual trend and one possible source of this is because the correction we apply is perturbative and so higher order terms could explain the residual. Another possible issue is that our model assumed infinitely small pixels and we leave an exploration of this assumption to future work.

\begin{figure}
\subfloat[Different seeing conditions]{
\centering
\includegraphics[width=.50\textwidth]{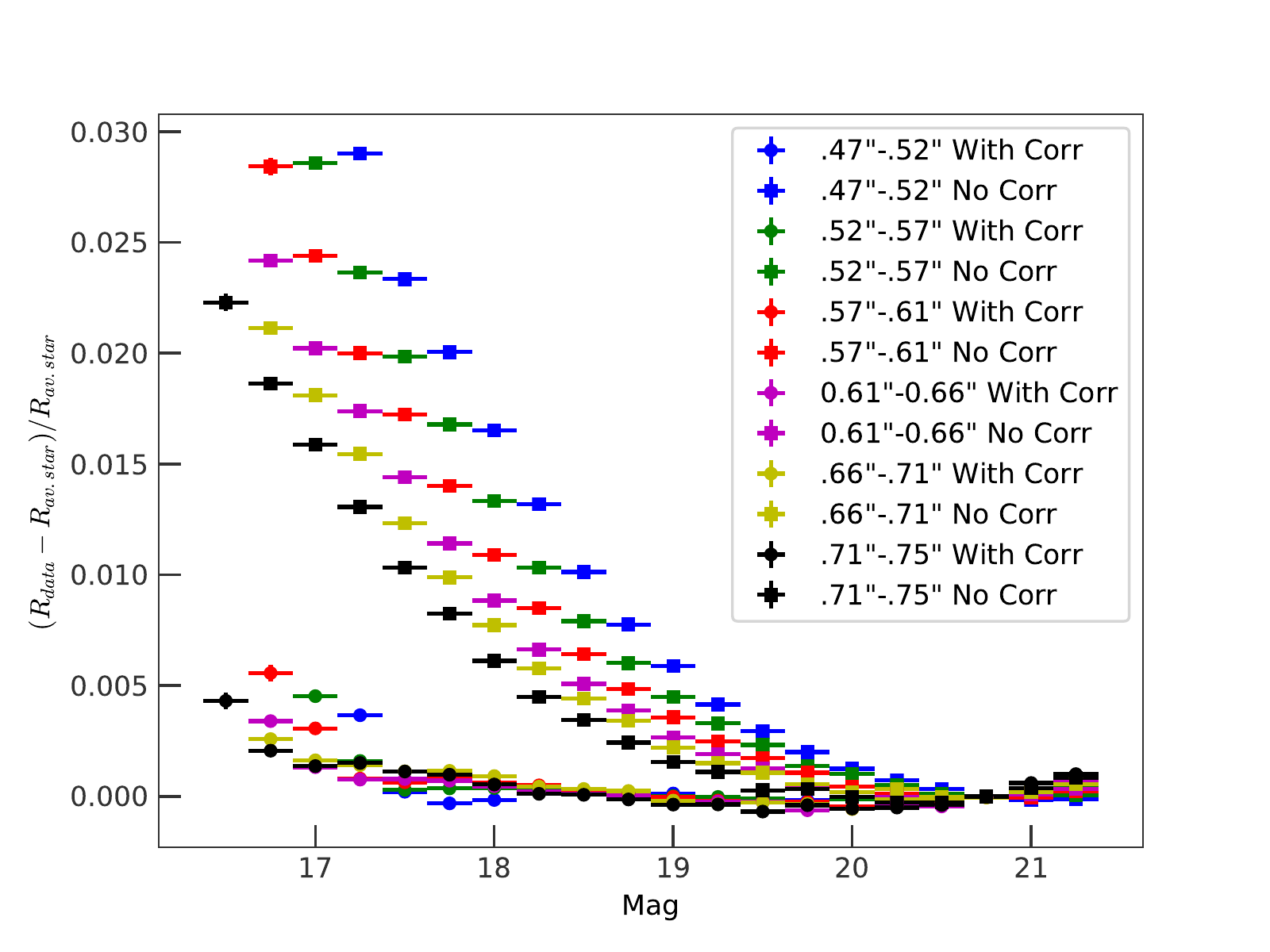}
\label{fig:avStarRadDifferentSeeing}
}\qquad
\subfloat[Different observation bands]{
  \centering
\includegraphics[width=.50\textwidth]{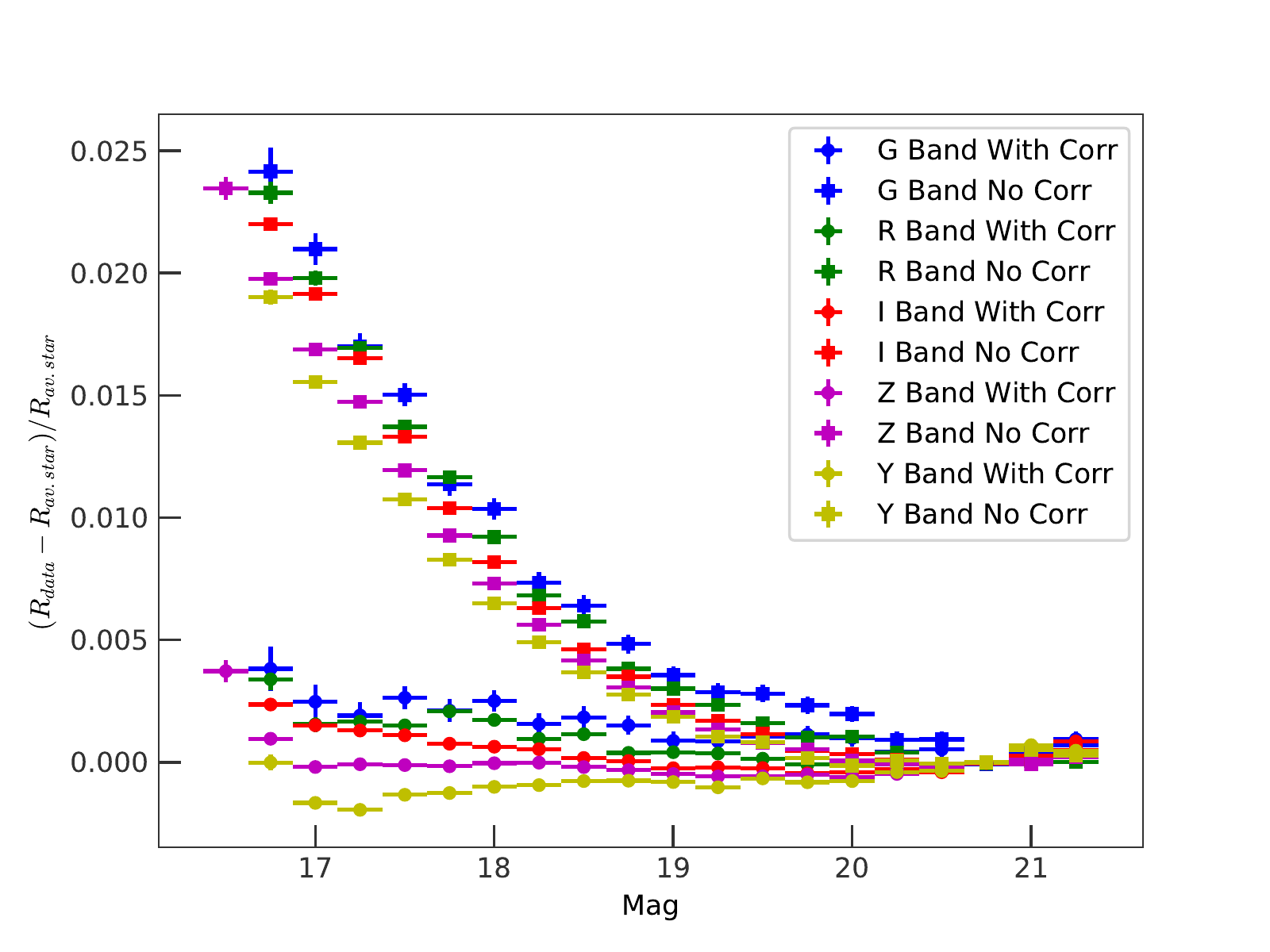}
\label{fig:avStarRadDifferentBands}
}
\caption{We investigate how our method performs for different seeing conditions, figure (a), and for different wavelengths, figure (b). For both cases we plot the different between the measured stars radius and the average star, as described in the text. We find that our correction's performance is largely independent of the seeing, removing the luminosity trend across a large range of magnitudes for all seeing conditions. When we investigate the wavelength dependence we use only stars with $0.517"<$ FWHM$<0.565"$. We find a weak wavelength dependence, with redder wavelengths (\textit{y},\textit{z} bands) showing a weaker effect than the bluer bands (\textit{g}, \textit{r}). In this work we used a wavelength independent kernel but these results motivate the use of different kernels for different wavelengths.}
\end{figure}

In figure \ref{fig:avStarRadDifferentSeeing} we show the size residuals when considering a range of different seeing conditions for stars observed in the \textit{i} band. The broadening by the brighter fatter effect is stronger for stars observed in conditions with smaller seeing as the stars are more concentrated with steeper profiles. It can be seen that our correction performs almost equally well across the range of seeing conditions. Figure \ref{fig:avStarRadDifferentBands} shows the residuals when applying the correction to the \textit{grizy} filters for stars observed when $0.52"<$ FWHM$<0.57"$. Here we can see a weak trend with frequency, with longer wavelengths less effected by the brighter fatter effect than the shorter wavelengths. This wavelength dependence arises as the different wavelength photons generate electron-hole pairs, on average, at different depths in the depleted region of the CCD. In this work we used a correction kernel obtained from flat fields from all bands; the wavelength dependence shown here demonstrates the necessity of using different kernels at different wavelengths.

\subsection{PSF Model Method}
In this method we compare our stars to a luminosity independent PSF model. In this work, we use PSFEx \citep{Bertin2011} to model the PSF across the camera.  We use a modified version of PSFEx that has been adapted to run in the HSC pipeline whilst leaving the core of the algorithm unchanged \citep{HSCpipeline}.  Potential stars are chosen by looking for a cluster of objects with the same size.  We extract cutout images of 41$\times$41 pixels around each candidate.  These postage stamps are fed directly into PSFEx.  Each star is modeled using pixelized images, sampled at the native pixel scale.  This is set with the PSFEx option $ \texttt{BASIS\_TYPE=PIXEL\_AUTO}$. Along with this, we set $\texttt{BASIS\_NUMBER}=20$, indicating that the full modeling is only done for the 400 brightest pixels.  The flux model for each star is normalized using a 12 pixel radius aperture magnitude.  We fit the PSF model to each CCD independently, using a second order polynomial in $x$ and $y$ to interpolate between stars.

\begin{figure}
\subfloat[Radii residuals of stars]{
  \centering
  \includegraphics[width=.50\textwidth]{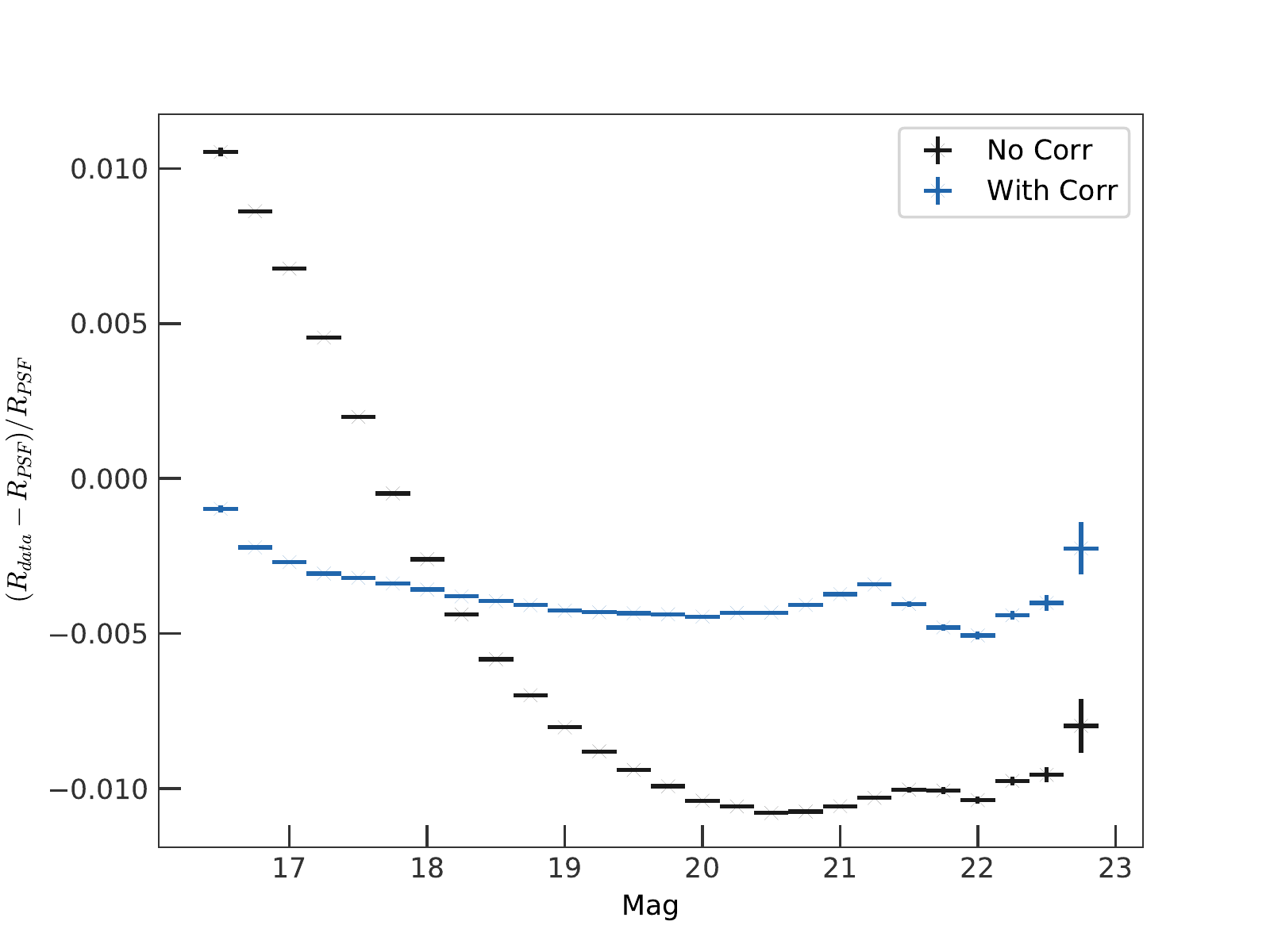}
 \label{fig:starRadii_PSF}
}\qquad
\subfloat[Two dimensional residuals of stacked stars]{  \centering
  \includegraphics[width=.50\textwidth]{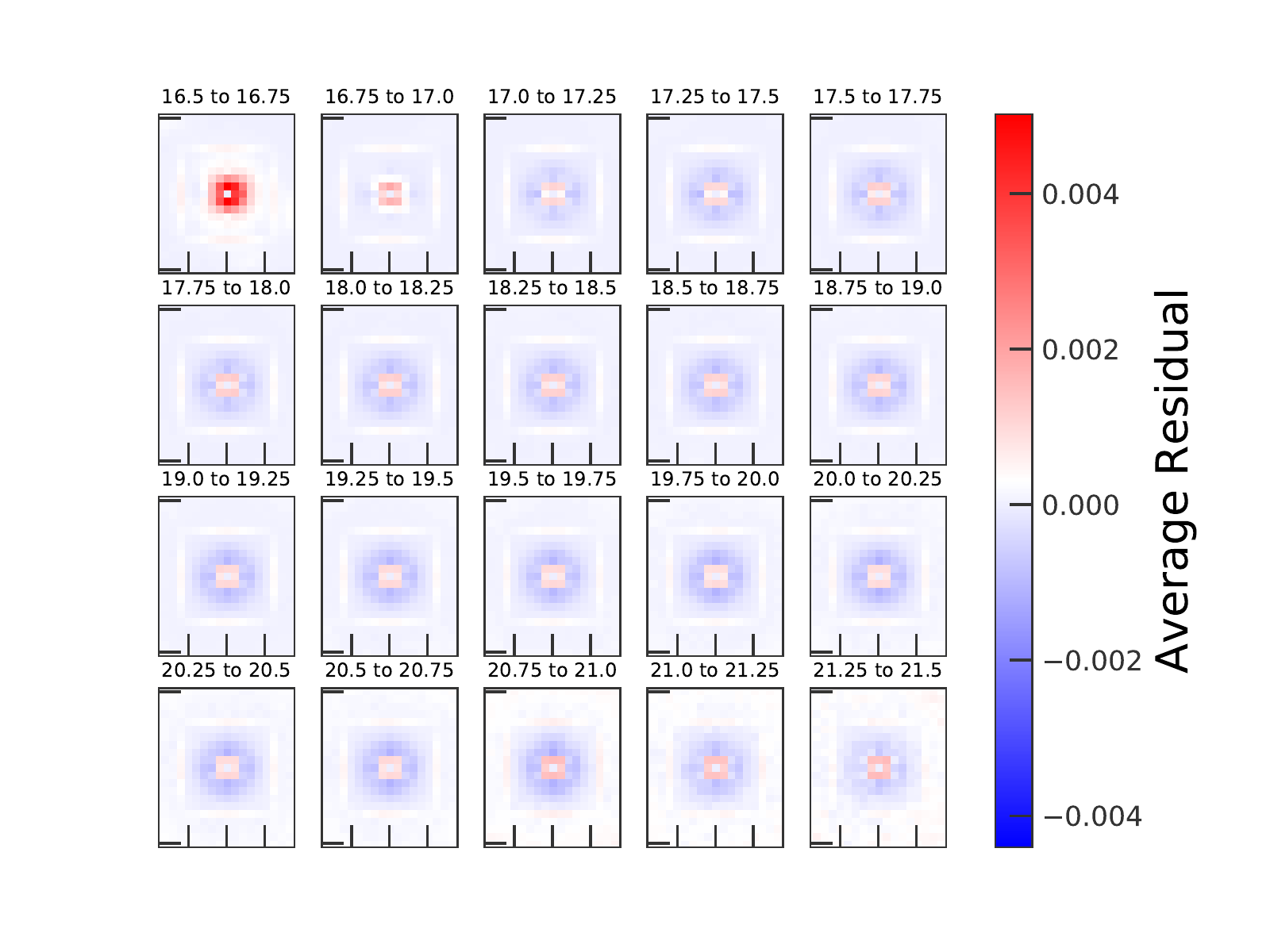}
 \label{fig:starRadii2D_PSF}
}
  \caption{In figure (a) we see the stellar radius residuals when comparing to the PSF model. It can be seen that our correction removes most of the luminosity dependence of the effect, however a residual offset remains between the stars and the PSF model. We investigate this further in figure (b) by viewing the two dimensional residuals between the star, divided by its peak flux, and the PSF model, also scaled by its peak value, and binned by instrumental flux. The brightest stars are in the top left and the faintest stars are in the bottom right. The resulting residual pattern is largely independent of luminosity as is the source of the offset in the stellar radii; the pattern is thought to arise due to a PSF modeling issue. }

\end{figure}

\begin{figure}
\subfloat[$e_1$ ellipticity]{
\centering
\includegraphics[width=.50\textwidth]{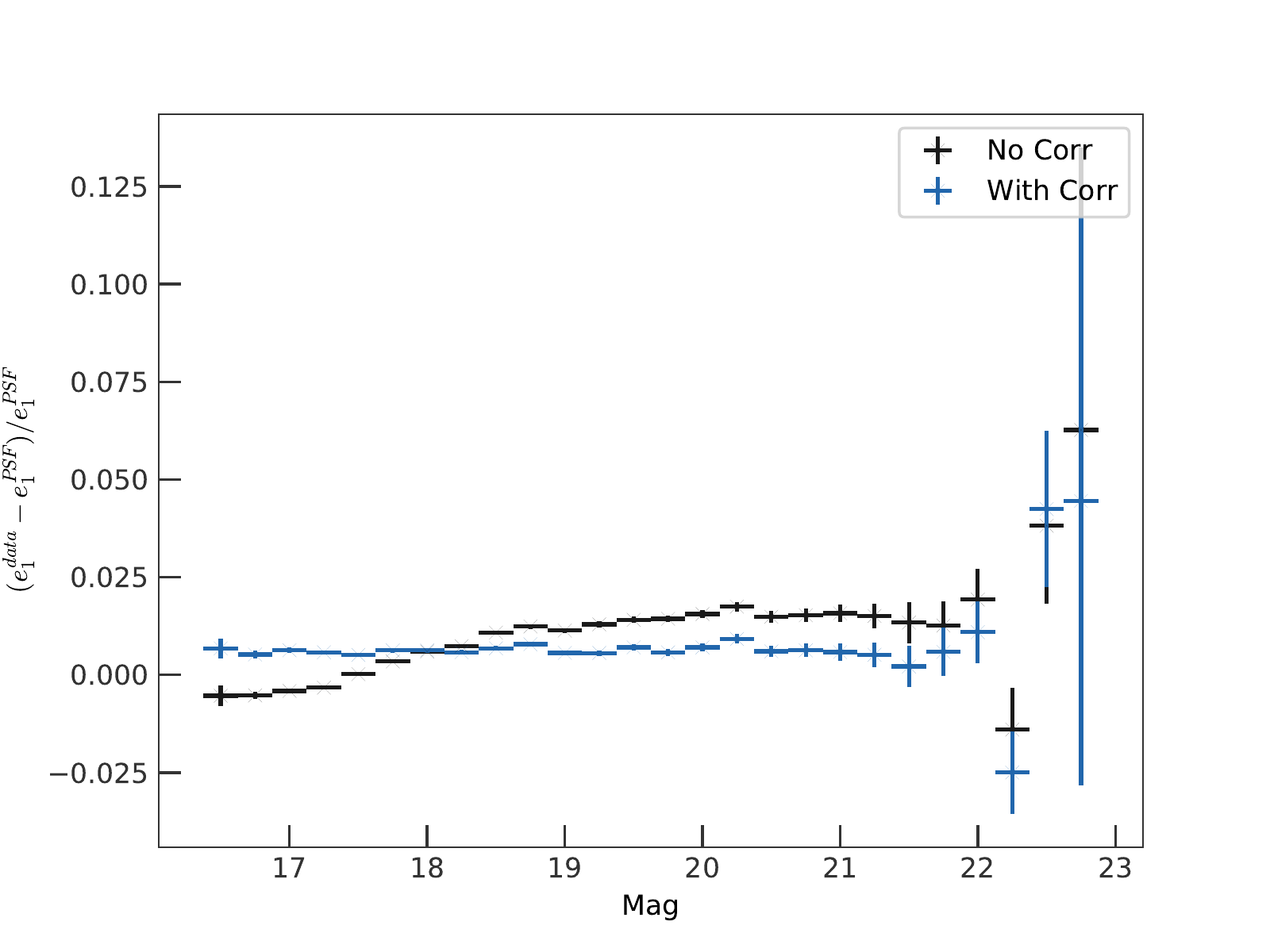}
\label{fig:e1Moments_PSF}
}\qquad
\subfloat[$e_2$ ellipticity]{
  \centering
\includegraphics[width=.50\textwidth]{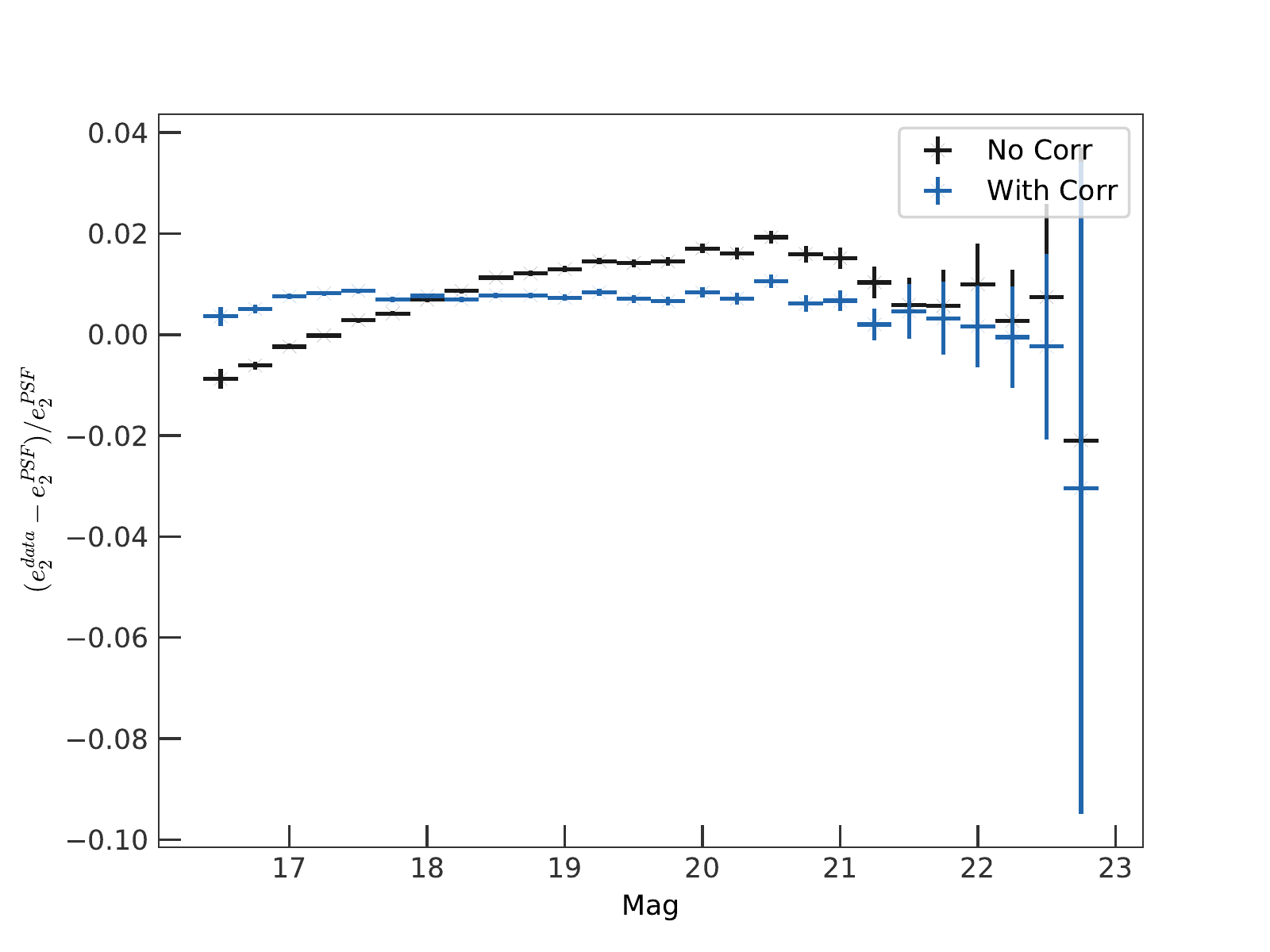}
\label{fig:e2Moments_PSF}
}
\caption{ The shape dependence of the stars with brightness is explored with the two ellipticity measures described in the text. We plot the fractional difference between the two ellipticiticies of stars and the PSF model. It can be seen that our correction does alter the shapes of the objects and leaves an almost luminosity independent residual. As in the case of the radii residuals we see a residual shape offset between the stars and PSF model.}
\label{fig:eMoments_PSF}
\end{figure}

In  figure  \ref{fig:starRadii_PSF} and  figure \ref{fig:eMoments_PSF} we  show the radius and ellipticity residuals after applying our correction when using the PSF model and again we find that the trend is almost completely removed. It can also be seen that there remains a difference between the stars and the PSF model (hence the residuals are not centered on zero); this offset is indicative of an issue with our PSF model (as this is not seen in the average star case). To further examine this in figure \ref{fig:starRadii2D_PSF} we plot the 2D residuals of the star minus the PSF scaled to the peak of the star at that point and subtracted. If the PSF model were perfect we would expect to see residuals consistent with noise. Instead we see a distinctive residual pattern showing that the PSF model is systematically different from the PSF. This residual pattern is almost independent of luminosity indicating that the PSF model fits all the different luminosity star's equally well. The offset between the PSF model and stars with PSFx has been seen by several different groups \citep{Kuijken2015,Jarvis2016}. We also find that the offset varies as a function of seeing; this implies that there is an issue with PSFEx as function of object width, with narrower objects being more severely affected. Once this offset is subtracted the residuals are consistent with those from the average star method, as can be seen in figure \ref{fig:futureRequirements}.

 It is also interesting to see that the residuals, especially in the uncorrected case,  have a difference shape to the average star case - for example in the residuals for the radius, the difference at the bright end is reduced in PSF model case with respect to the average star residual difference. It is thought that this arises as the PSF model is attempting to fit part of the brighter fatter effect as a spacial variation.
 
\section{Weak Lensing Survey Limits}
\label{ch:lensingReq}
In on going and upcoming weak lensing surveys the control of systematics is crucial to preform the desired science. Here we compare the results from this algorithm to the weak lensing requirements for HSC and LSST.  In making these comparisons, we have subtracted off the mean offset, as we expect this is due to problems in PSF modeling.  We use the requirements on the PSF size as outlined in \citet{HSCfirstyearShear}.  These requirements are derived by setting the systematic error to be less than half the statistical error in the first year HSC data from science analyses of galaxy-galaxy lensing and cosmic shear.  It uses the formalism derived in \citet{Hirata2004} to propagate the uncertainty in PSF size into a multiplicative shear bias. To extrapolate the comparisons to the final HSC and LSST requirements we have simply scaled the first year HSC requirements by the increase in area \citep{LSST2012}.  

\begin{figure}
\centering
\includegraphics[width=.50\textwidth]{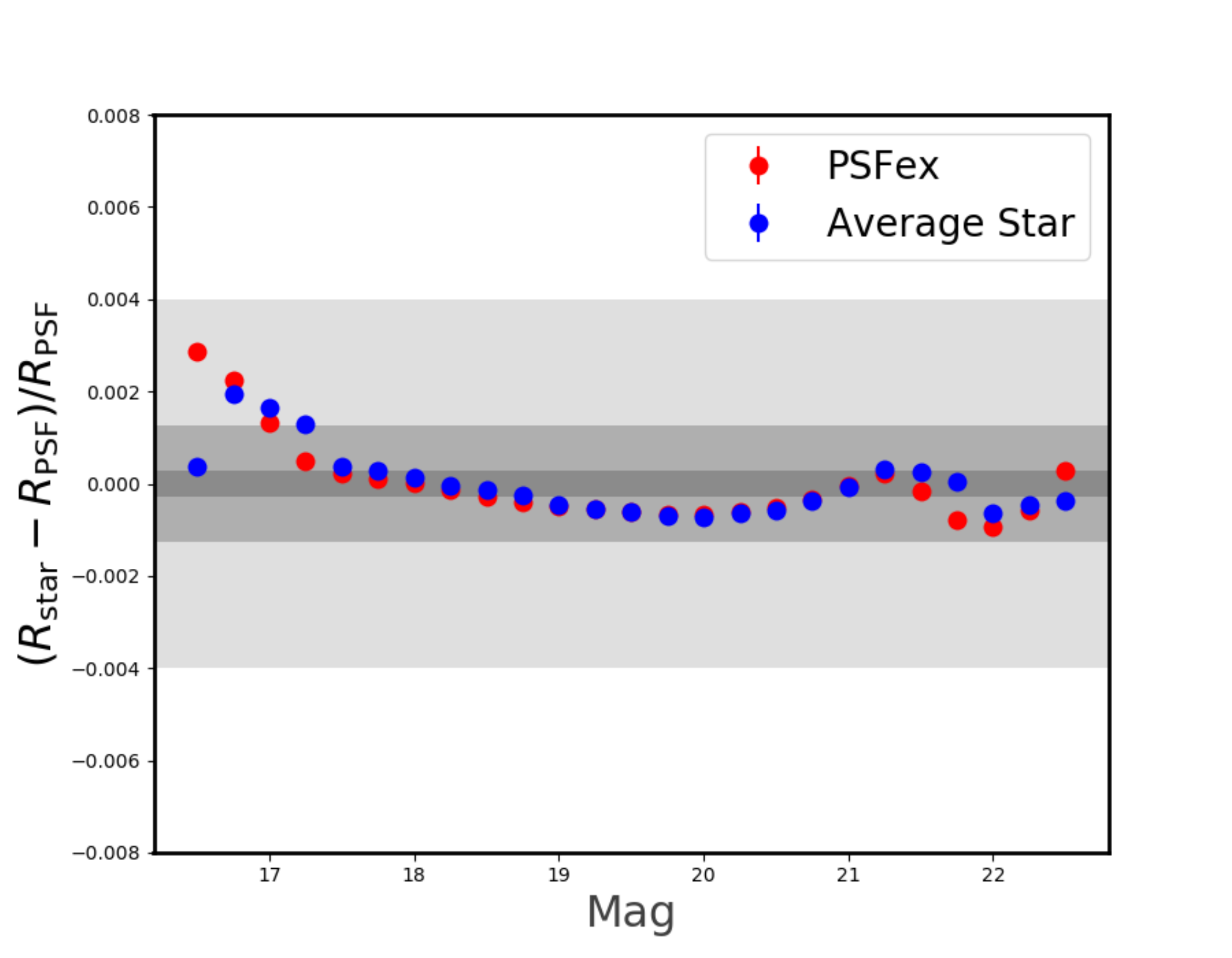}
\caption{ The PSF residuals after applying our brighter fatter correction  are compared to the requirements of the HSC survey (both first year and five year) and the LSST experiment. We see that our correction should be  sufficient over a wide rage of magnitudes for the HSC experiment but improvements must be made for LSST.}
\label{fig:futureRequirements}
\end{figure}

There are a number of reasons why this simple area scaling is not correct, including: the galaxy populations potentially being different and thus changing how the size bias propagates to shear bias, some of the lensing covariance terms do not scale with area, and future surveys will have more contiguous fields which will lower the errors at large scales.  Because of these factors, the comparison with LSST and Final HSC data are only meant to be a rough estimate.  Figure \ref{fig:futureRequirements} shows the bias in the measured size as a function of magnitude with the derived requirements.  We can see that we meet the requirements for the Final HSC dataset, but that we need to improve our correction by a factor of two or more in order  to reach the required levels for LSST.

\section{Conclusion}

We discussed and implemented a new model that attempts to correct the brighter fatter effect and it is seen that this model offers a reasonable correction for stars and flat fields. We find that our correction performs well across a wide range of stellar luminosities and seeing conditions. We see a weak wavelength dependence similar to that seen by the Dark Energy Survey \citep{DESbrighterFatter} and this reinforces the need for using different kernels at different wavelengths.  Further work is required to investigate whether the accuracy of our correction can be improved, perhaps by taking into account the finite pixel size in our theory or further understanding higher order corrections, so that the brighter-fatter effect is not a limiting systematic for future surveys, such as LSST.

\section{Acknowledgements}

WRC acknowledges support from the WFIRST project. The Hyper Suprime-Cam (HSC) collaboration includes the astronomical
communities of Japan and Taiwan, and Princeton University. The HSC
instrumentation and software were developed by the National Astronomical
Observatory of Japan (NAOJ), the Kavli Institute for the Physics and
Mathematics of the Universe (Kavli IPMU), the University of Tokyo, the
High Energy Accelerator Research Organization (KEK), the Academia Sinica
Institute for Astronomy and Astrophysics in Taiwan (ASIAA), and
Princeton University. Funding was contributed by the FIRST program from
Japanese Cabinet Office, the Ministry of Education, Culture, Sports,
Science and Technology (MEXT), the Japan Society for the Promotion of
Science (JSPS), Japan Science and Technology Agency (JST), the Toray
Science Foundation, NAOJ, Kavli IPMU, KEK, ASIAA, and Princeton
University. HM is supported by the Jet Propulsion Laboratory, California
Institute of Technology, under a contract with the National Aeronautics
and Space Administration.
This paper makes use of software developed for the Large Synoptic Survey
Telescope. We thank the LSST Project for making their code available as
free software at \url{http://dm.lsst.org}

The Pan-STARRS1 Surveys (PS1) have been made possible through
contributions of the Institute for Astronomy, the University of Hawaii,
the Pan-STARRS Project Office, the Max-Planck Society and its
participating institutes, the Max Planck Institute for Astronomy,
Heidelberg and the Max Planck Institute for Extraterrestrial Physics,
Garching, The Johns Hopkins University, Durham University, the
University of Edinburgh, Queen's University Belfast, the
Harvard-Smithsonian Center for Astrophysics, the Las Cumbres Observatory
Global Telescope Network Incorporated, the National Central University
of Taiwan, the Space Telescope Science Institute, the National
Aeronautics and Space Administration under Grant No. NNX08AR22G issued
through the Planetary Science Division of the NASA Science Mission
Directorate, the National Science Foundation under Grant
No. AST-1238877, the University of Maryland, and Eotvos Lorand
University (ELTE) and the Los Alamos National Laboratory.

Based on data collected at the Subaru Telescope and retrieved
from the HSC data archive system, which is operated by Subaru Telescope
and Astronomy Data Center at National Astronomical Observatory of Japan.

This work is supported in part by JSPS KAKENHI (Grant Number
JP~15H03654) as well as MEXT Grant-in-Aid for Scientific Research on
Innovative Areas (15H05887, 15H05892, 15H05893, 15K21733).

\bibliographystyle{aasjournal}
\bibliography{Project}
\end{document}